\documentclass{aoscm}
\usepackage[dvips]{color}
\usepackage{graphics,graphicx}
\usepackage{latexsym}
\usepackage{times}
\usepackage{cite}
\usepackage{natbib}  
\usepackage{amsbsy} 
\def\newblock{\hskip .11em plus .33em minus .07em}
\newtheorem{theorem}{Theorem}[section]   
   
\newtheorem{remark}{Remark}[section]   
\newtheorem{lemma}{Lemma}[section]

\begin{document}

\SPECFNSYMBOL{}{}{}{}{}{}{}{}{}

\AOSMAKETITLE

\AOSAMS{Primary 62H05, 62G35. Secondary 62-09, G2H25, G2H30.}

\AOSKeywords{Affine equivariance, asymptotic theory, breakdown point, eigenprojections, elliptical distributions, principal components, relative efficiency, robustness, 
spatial sign covariance matrix, Tyler's scatter matrix.}

\AOStitle{THE ASYMPTOTIC INADMISSIBILITY OF THE SPATIAL SIGN COVARIANCE MATRIX \\ FOR ELLIPTICALLY SYMMETRIC DISTRIBUTIONS}

\AOSauthor{Andrew F.\ Magyar  and David E.\ Tyler \thanks{\emph{Acknowledgements.} The research for both authors was supported in part by NSF Grant DMS-0906773.}}

\AOSaffil{The United States Census Bureau and Rutgers University}

\AOSlrh{A.F.\ MAGYAR AND D.E.\ TYLER}

\AOSrrh{INADMISSIBILITY OF TE SPATIAL SIGN COVARIANCE MATRIX}

\AOSAbstract{\begin{center} \textbf{ABSTRACT} \end{center}
The asymptotic efficiency of the spatial sign covariance matrix (\emph{SSCM}) relative to affine equivariant estimates of scatter is studied in detail.
In particular, the \emph{SSCM} is shown to be asymptoticaly inadmissible, i.e.\  the asymptotic variance-covariance matrix of the consistency corrected
\emph{SSCM} is uniformly smaller than that of its affine equivariant counterpart, namely Tyler's scatter matrix.  Although the \emph{SSCM}
has often been recommended when one is interested in principal components analysis, the degree of the
inefficiency of the \emph{SSCM} is shown to be most severe in situations where principal components are of most interest. A finite
sample simulation shows the inefficiency of the \emph{SSCM} also holds for small sample sizes, and that the
asymptotic relative efficiency is a good approximation to the finite sample efficiency for relatively modest sample sizes.
}

\maketitle

\BACKTONORMALFOOTNOTE{3}

\renewcommand\arraystretch{1.05}
\newcommand{\bmu}{\boldsymbol{\mu}}
\newcommand{\btheta}{\boldsymbol{\theta}}
\newcommand{\bEta}{\boldsymbol{\eta}}
\newcommand{\bA}{\boldsymbol{A}}
\newcommand{\ba}{\boldsymbol{a}}
\newcommand{\bb}{\boldsymbol{b}}
\newcommand{\bee}{\boldsymbol{e}}
\newcommand{\bq}{\boldsymbol{q}}
\newcommand{\bt}{\boldsymbol{t}}
\newcommand{\bu}{\boldsymbol{u}}
\newcommand{\bx}{\boldsymbol{x}}
\newcommand{\by}{\boldsymbol{y}}
\newcommand{\bz}{\boldsymbol{z}}
\newcommand{\bB}{\boldsymbol{B}}
\newcommand{\bC}{\boldsymbol{C}}
\newcommand{\bD}{\boldsymbol{D}}
\newcommand{\bE}{\boldsymbol{E}}
\newcommand{\bI}{\boldsymbol{I}}
\newcommand{\bJ}{\boldsymbol{J}}
\newcommand{\bM}{\boldsymbol{M}}
\newcommand{\bN}{\boldsymbol{N}}
\newcommand{\bP}{\boldsymbol{P}}
\newcommand{\bQ}{\boldsymbol{Q}}
\newcommand{\bR}{\boldsymbol{R}}
\newcommand{\bS}{\boldsymbol{S}}
\newcommand{\bT}{\boldsymbol{T}}
\newcommand{\bV}{\boldsymbol{V}}
\newcommand{\bW}{\boldsymbol{W}}
\newcommand{\bX}{\boldsymbol{X}}
\newcommand{\bY}{\boldsymbol{Y}}
\newcommand{\bZ}{\boldsymbol{Z}}
\newcommand{\bzero}{\boldsymbol{0}}
\newcommand{\bDelta}{\boldsymbol{\Delta}}
\newcommand{\bLambda}{\boldsymbol{\Lambda}}
\newcommand{\bSigma}{\boldsymbol{\Sigma}}
\newcommand{\bGamma}{\boldsymbol{\Gamma}}
\newcommand{\mA}{\mathcal{A}}
\newcommand{\mE}{\mathcal{E}_d}
\newcommand{\mI}{\mathcal{I}}
\newcommand{\mV}{\boldsymbol{\mathcal{V}}}
\newcommand{\mD}{\mathcal{D}}
\newcommand{\mH}{\mathcal{H}}
\newcommand{\mK}{\mathcal{K}}
\newcommand{\mM}{\mathcal{M}}
\newcommand{\mN}{\mathcal{N}}
\newcommand{\mP}{\mathcal{P}}
\newcommand{\mS}{\mathcal{S}}
\newcommand{\mT}{\mathcal{T}}
\newcommand{\bXi}{\boldsymbol{\Xi}}
\newcommand{\bhP}{\widehat{\boldsymbol{P}}}
\newcommand{\bhS}{\widehat{\boldsymbol{\mathcal{S}}}}
\newcommand{\bhT}{\widehat{\boldsymbol{\mathcal{T}}}}
\newcommand{\bhmu}{\widehat{\boldsymbol{\mu}}_n}
\newcommand{\red}{\textcolor[rgb]{.7,0,0}}
\newcommand{\trace}{\mbox{trace}}

\section{Introduction}
Multivariate procedures are implemented to help understand the relationships between several quantitative variables of interest.
The most commonly used methods rely heavily on the sample variance-covariance matrix. It is now well known 
that the sample variance-covariance matrix is highly non-robust, being extremely sensitive to outliers and being very inefficient at longer 
tailed distributions. Consequently, there have been many proposed robust alternatives to the sample variance-covariance matrix, such as the $M$-estimates 
\citep{Maronna76,Huber77}, the minimum volume ellipsoid and the minimum covariance determinant estimates \citep{Rousseeuw85},
the Stahel-Donoho estimates \citep{Stahel81,Donoho82}, the $S$-estimates \citep{Davies87,Lopuhaa89}, the P-estimates \citep{Maronna92,Tyler94}, $CM$-estimates \citep{Kent96}
and the $MM$-estimates \citep{Tatsuoka00,Tyler02}. All of these estimates of scatter are affine equivariant, and except for the $M$-estimates they
all have high breakdown points.  On the other hand, the computations of $M$-estimates are relatively easy and can be done via a simple IRLS algorithm,
whereas the high breakdown point scatter estimates are computationally intensive, especially for large samples and/or high dimensional data sets,
and are usually computed via approximate or probabilistic algorithms. 

Due to the computational complexity of high breakdown point affine equivariant estimates of multivariate scatter, there has been recent interest in
high breakdown point estimates of scatter which are not affine equivariant but are computationally easy. One such estimate is the spatial sign covariance
matrix, hereafter referred to as \emph{SSCM}. The earliest often cited references to this estimate is \cite{Locantore99}, though the name \emph{SSCM} was introduced in
\cite{Visuri00}. The \emph{SSCM} is often recommended as a fast and easy high breakdown point method; see e.g.\ \cite{Locantore99} and the recent book by \cite{Maronna06}.  
In particular, since it is orthogonally equivariant, the \emph{SSCM} has been recommended in cases where only orthogonal equivariance and not full affine equivariance
is needed, such as in principal components analysis.

The lack of affine equivariance of the \emph{SSCM} arises since it is defined by down-weighing observations based upon their Euclidean distances from an estimated center 
of the data set. When the data is presumed to arise from a multivariate normal distribution or more generally from an elliptically symmetric distribution, one might 
conjecture that the \emph{SSCM} would be relatively inefficient whenever the elliptical distribution is far from spherical, i.e. when the components of the multivariate vector 
are highly correlated. This can be problematic since one usually implements multivariate procedures in situations where one suspects strong correlations 
between the variables, as is the case when one does a principal components analysis. Our goal here is then to study the efficiency of the \emph{SSCM} under general covariance 
structures. 

An important property of the \emph{SSCM} is that its asymptotic distribution does not depend on the particular elliptical family being sampled.  This is also true for the finite 
sample distribution of the \emph{SSCM} when the center of symmetry of the elliptical distribution is known.  An affine equivariant estimate possessing this same property is the 
distribution free M-estimate of scatter proposed by \cite{Tyler87a}, commonly referred to as Tyler's scatter matrix. This scatter matrix can be considered an affine equivariant 
version of the \emph{SSCM} and so for our purposes here we abbreviate it \emph{ASSCM}. The asymptotic relative efficiency (\emph{ARE}) of the \emph{SSCM} to 
an affine equivariant estimate of scatter, say $\widehat{\bGamma}$, can then be factored into the \emph{ARE} of the \emph{SSCM} to the \emph{ASSCM} and the \emph{ARE} of the \emph{ASSCM} to 
$\widehat{\bGamma}$. For an elliptically symmetric distribution, the first factor depends only on the underlying covariance structure and not on the particular elliptical family, whereas the
second factor depends only on the particular elliptical family and not on the underlying covariance structure. Hence, the \emph{ASSCM} serves as a convenient benchmark for 
understanding the effect the lack of affine equivariance has on the efficiency of \emph{SSCM} for varying covariance structures.  

In section \ref{Theory}, we show that our conjecture regarding the inefficiency of the \emph{SSCM} is indeed true.  It is shown the 
performance of the \emph{ASSCM} dominates that of the \emph{SSCM} in that the \emph{ARE} of the \emph{SSCM} to the \emph{ASSCM} is never greater than one 
regardless of the underlying covariance structure and can approach zero when the component variables are highly correlated.  The form of the \emph{ARE} is complicated,
but in special cases it can be expressed in terms of the Gauss hypergeometric functions. A simulation study, presented in section \ref{Sim}, shows that the advantage 
of the \emph{ASSCM} over the \emph{SSCM} holds for relatively small sample sizes and that the asymptotic results are quit accurate even for modest sample sizes. 
Some technical results and proof are reserved for section \ref{Appendix}, an appendix. The following section gives some needed background information on elliptical distributions,
affine equivariance, the spatial sign covariance matrix and Tyler's scatter matrix, as well as sets up the notation for the paper. 

\section{Preliminaries} \label{Prelim}
\subsection{Elliptical distributions and equivariance} \label{subell}

Elliptically symmetric distributions provide a simple generalization of the multivariate normal distribution and 
are often used to ascertain how multivariate statistical methods perform outside of the normal family.  An elliptically symmetric 
distribution in $\mathbb{R}^d$ is defined to be one arising from an affine transformation of a spherically symmetric distribution, i.e.\ if $\bz \sim_\mD \bQ\bz$ for 
any $d \times d$ orthogonal matrix $\bQ$, then the distribution of $\bx = \bA\bz + \bmu$ is said to have an elliptically symmetric 
distribution with center $\bmu \in \mathbb{R}^d$ and scatter matrix $\bGamma = \bA\bA^T$, see e.g.\ section 13.2 in \cite{Bilodeau99}. The distribution of $\bx$ is 
characterized by $\bmu$, $\bGamma$ and the distribution of $\bz$. The distribution of $\bz$ itself can be characterized as $\bz \sim_\mD R_G \bu_d$,
with $\bu_d$ having a uniform distribution on the unit $d$-dimensional sphere independent of its radial component $R_G $, 
a non-negative random variable with distribution function $G$. In particular, $(\bu_d, R_G) \sim_\mD (\bz/\|\bz\|, \|\bz\|)$, where the norm 
refers to the usual Euclidean norm in $\mathbb{R}^d$, i.e.\ $\|\bz\| = \sqrt{\bz^{T}\bz}$.
Thus, we denote the distribution of 
\begin{equation} \label{Ed}
\bx \sim_\mD R_G \bGamma^{1/2} \bu_d + \bmu
\end{equation} 
by $\mE(\bmu,\bGamma;G)$, where $\bGamma^{1/2}$ refers to the unique positive definite square root of $\bGamma$. If the distribution of $\bx$ is also 
absolutely continuous, then like the multivariate normal distribution, its density has concentric elliptical contours, with a density 
of the form
\begin{equation} \label{ellip}
f\left(\bx;\boldsymbol{\mu},\bGamma,g\right)=det\left(\bGamma\right)^{-\frac{1}{2}}g\left(\left(\bx-\boldsymbol{\mu}\right)^{T}\bGamma^{-1}\left(\bx-\boldsymbol{\mu}\right)\right),
\end{equation}
for $\bx \in \mathbb{R}^d$, where $g$ is some non-negative function and $\bGamma \in \mathcal{P}_d$, the class of $d \times d$ symmetric positive definite
matrices. The relationship between $g$ and $G$ is $G'(r) = a_d r^{d-1} g(r^2)$ with the constant $a_d = 2 \pi^{d/2}/\Gamma\left(d/2\right)$, where the non-bold
$\Gamma$ refers to the usual gamma function. 

Note that the scatter parameter $\bGamma$ is only well defined up to a scalar multiple, i.e.\ if $\bGamma$ satisfies the definition of a 
scatter matrix for a given elliptically symmetric distribution, then $\lambda \bGamma$ also does for any $\lambda > 0$. If
no restrictions are placed on the function $G$, then the parameter $\bGamma$ is confounded with $G$. If $\bx$ possesses first moments, 
then $\bmu = E[\bx]$, and if it possesses second moments, then $\bGamma \propto \bSigma$, the variance-covariance matrix, or more specifically 
$\bSigma = \lambda(G) \bGamma$ with $\lambda(G) = E[R_G^2]/d$ . 

The parameters of an elliptically symmetric distribution are affinely equivariant. That is, if $\bx \sim_\mD \mE(\bmu,\bGamma;G)$, then 
$\bB\bx + \bb \sim_\mD \mE(\bB\bmu +\bb, \bB\bGamma \bB^{T};G)$ for any non-singular $d \times d$ matrix $\bB$ and any vector $\bb \in \mathbb{R}^d$.  
Consequently, the maximum likelihood estimates for the parameters of an elliptically symmetric 
distribution based on a given function $G$ are affinely equivariant.  Given a $d$-dimensional sample $\bx_1,\ldots,\bx_n$, estimates of  $\bmu$ and $\bGamma$ 
are said to be affinely equivariant if an affine transformation of the data $\bx_i \rightarrow \bB\bx_i +\bb$ for $i = 1, \ldots, n$ induces on the estimates 
the transformations $\widehat{\bmu} \rightarrow \bB\widehat{\bmu} + \bb$ and $\widehat{\bGamma} \rightarrow \bB \widehat{\bGamma} \bB^{T}$.
In addition to the maximum likelihood estimates associated with elliptically symmetric distributions, which include the sample mean vector and the sample 
covariance matrix, many proposed robust estimates of multivariate location and scatter, such as the multivariate \emph{M}-estimates, are affinely equivariant. 

The covariance matrix, if it exists, of an affine equivariant estimate of scatter under random sampling from an $\mE(\bmu,\bGamma;G)$ distribution has a relatively
simple form, namely
\begin{equation} \label{varaff}
var(vec\{\widehat{\bGamma}\}) = \sigma_{1,n,\widehat{\bGamma},G} (\bI_{d^2} + \mK_{d,d}) (\bGamma \otimes \bGamma) + \sigma_{2,n,\widehat{\bGamma},G} 
vec(\bGamma)vec(\bGamma)^T.
\end{equation}
The notation $vec(\bA)$ refers to the $pq$-dimensional vector obtained from stacking the columns of the $p \times q$ dimensional matrix $\bA$, $\mathcal{K}_{d,d}$ refers to
the $d^2 \times d^2$ commutation matrix, and $\otimes$ refers to the Kronecker product. This notation is reviewed in Appendix \ref{matrix}. The terms $\sigma_{1,n,\widehat{\bGamma},G}$ and 
$\sigma_{2,n,\widehat{\bGamma},G}$ are scalars that depend on the sample size, the particular estimate $\widehat{\bGamma}$ and on the underlying family of 
elliptical distribution, i.e.\ on $G$, but not on $\bmu$ or $\bGamma$, see \cite{Tyler83} for more details. When focusing on the \emph{shape} of the scatter
parameter, i.e.\ on $\bGamma$ up to a scalar multiple, the second 
scalar becomes unimportant, at least asymptotically.  That is, if $\widehat{\bGamma}$ converges in probability and is asymptotically normal, then if the function $\mH$ is
such that $\mH(\bGamma) = \mH(\beta \bGamma)$ for any $\beta > 0$, e.g.\ $\mH(\bGamma) = vec(\bGamma)/\trace(\bGamma)$, then 
\begin{equation} \label{shape}
\sqrt{n}\{\mathcal{\mH}(\widehat{\bGamma}) - \mathcal{\mH}(\bGamma)\} \rightarrow_\mD Normal_{p}\left(\bzero, \sigma_{1,\widehat{\bGamma},G} \mM(\bGamma)\right),
\end{equation}
where $p$ is the dimension of $\mathcal{H}$, $\mathcal{M}$ is a function dependent on $\mathcal{H}$ with $\mM(\beta \bGamma) = \mM(\bGamma)$  for any
$\beta > 0$, and $\sigma_{1,\widehat{\bGamma},G}$ is again a scalar dependent only on the particular estimate and on $G$. Thus, the asymptotic relative
efficiency at $\mE(\bmu,\bGamma;G)$ of any shape component based on one affine equivariant asymptoptically normal estimate of scatter versus another
reduces to the scalar 
\begin{equation} \label{areaff}
ARE\left(\mH(\widehat{\bGamma}_1), \mH(\widehat{\bGamma}_2)\right) = \frac{\sigma_{1,\widehat{\bGamma}_2,G}}{\sigma_{1,\widehat{\bGamma}_1,G}}, 
\end{equation}
which is not dependent on the shape function $\mathcal{H}$ nor on the parameters $\bmu$ and $\bGamma$. We again refer the reader to \cite{Tyler83} for more details.

\subsection{Spatial Sign Covariance Matrix} \label{subSSCM}
Given a $d$-dimensional sample $\textbf{x}_{1}, \ldots, \textbf{x}_{n}$, the spatial sign covariance matrix, or \emph{SSCM}, about a point $\bmu$ is defined to be
\begin{equation} \label{sccm}
\bhS(\bmu) = \frac{1}{n} \sum^{n}_{i=1} \frac{\left( \bx_{i}-\bmu \right)\left(\bx_{i}-\bmu \right)^{T}}{\left(\bx_{i}-\bmu \right)^{T}
\left(\bx_{i}-\bmu\right)} = \frac{1}{n} \sum^{n}_{i=1} \btheta_i \btheta_i^{T},
\end{equation}
where $\btheta_i = (\bx_i-\bmu)/\|\bx_i-\bmu\|$.  Note that by definition, $\trace\{\bhS(\bmu)\} = 1$.
Under general random sampling $\boldsymbol{\Xi} = E\left[\bhS(\bmu)\right] = E\left[\btheta \btheta^{T}\right]$, where $\btheta = (\bx-\bmu)/\|\bx-\bmu\|$, 
and by the law of large numbers $\bhS(\bmu)$ is consistent for $\bXi$. Also, since $\|\btheta\| = 1$, the central limit theorem immediately gives
\begin{equation} \label{clt}
\sqrt{n} ~ vec\{ \bhS(\bmu) - \bXi \} \rightarrow_\mD Normal_{d^2}\left(\bzero, \mV_\mS \right),
\end{equation}
where $\mV_\mS = var\left(vec\{\btheta\btheta^{T}\}\right) = var\left(\btheta \otimes \btheta \right)  =
 \left(E[\btheta\btheta^{T} \otimes \btheta\btheta^{T}] - vec(\bXi)vec(\bXi)^{T} \right)$. 
In practice, $\bmu$ is usually estimated, say by $\widehat{\bmu}_n$. 
The asymptotic distribution of the \emph{SSCM} when the center is estimated, that is for $\bhS(\widehat{\bmu}_n)$, which we hereafter abbreviate as $\bhS_n$ 
has only recently been obtained, see \cite{Durre13}. As the following lemma shows, under mild conditions the asymptotic distribution of 
$\bhS_n$ is the same as that of $\bhS(\bmu)$. Since the definition of $\bhS_n$ depends on the directions of $\bx_i - \widehat{\bmu}_n$, it is not 
surprising to note that the distribution of $\bx_i$ cannot be too concentrated about $\widehat{\bmu}_n$. 

\begin{lemma} \label{lemma-asym}
Suppose $\bx \in \mathbb{R}^k$ is a continuous random vector whose distribution is symmetric about $\bmu$, i.e. $(\bx -\bmu) \sim_\mD - (\bx -\bmu)$, and such that 
$E[\| \bx - \bmu \|^{-3/2}] < \infty$. For a $k$-dimensional random sample $\bx_1, \ldots, \bx_n$ of the random vector $\bx$, suppose
$\sqrt{n}(\widehat{\bmu}_n - \bmu) = O_\mP(1)$, then \mbox{$\sqrt{n}\{\bhS_n-\bhS(\bmu)\} \rightarrow_\mP \bzero$}.
\end{lemma}

Under random sampling from an $\mE(\bmu,\bGamma;G)$ distribution, the finite sample distribution of $\bhS(\bmu)$, and hence its asymptotic distribution as well as
the asymptotic distribution of $\bhS_n$, does not depend on the radial component $G$ and depends only on $\bGamma$ up to a scalar multiple. This follows since if 
$\bx \sim_\mD \mE(\bmu,\bGamma;G)$ then $\btheta \sim_\mD ACG_d(\bGamma)$, an angular central Gaussian distribution with parameter $\bGamma$, which does not 
depend on $G$, see e.g.\ \cite{Watson83} or \cite{Tyler87b} and furthermore, the $ACG_d(\bGamma) \sim_\mD ACG_d(\beta \bGamma)$ for any $\beta > 0$.
It is known that $\bXi$ is not proportional to $\bGamma$ and so in contrast to affine equivariant estimates of scatter, the \emph{SSCM} is not consistent 
for some multiple of $\bGamma$, see e.g. \cite{Boente99}. Consequently, a shape component of the \emph{SSCM} is not necessarily consistent for the 
corresponding shape component of $\bGamma$. A transformation of $\bhS_n$, though, can be made so that it is consistent for some multiple of $\bGamma$. This
is discussed in more detail in section \ref{Theory}. 

Although not affine equivariant, $\bhS_n$ is orthogonally equivariant provided the estimate of location $\widehat{\bmu}_n$ is orthogonally equivariant.  That is
if for any orthogonal matrix $\bQ$ and any vector $\bb \in \mathbb{R}^d$ the transformation $\bx_i \rightarrow \bQ \bx_i + \bb$ for $i = 1, \ldots n$
induces the transformation $\widehat{\bmu}_n \rightarrow \bQ\widehat{\bmu}_n + \bb$, then it also induces the transformation $\bhS_n \rightarrow \bQ\bhS_n\bQ^{T}$.
As mentioned in the introduction, the \emph{SSCM} has a high breakdown point. 

\subsection{Tyler's Scatter Matrix} \label{subTyler}
For the multivariate sample $\bx_{1},\ldots, \bx_{n}$,  \cite {Tyler87a} introduced the distribution-free M-estimate of multivariate scatter about a point
$\bmu$ as a solution to the implicit equation
\begin{equation} \label{tyler}
\bhT(\bmu) = \frac{d}{n} \sum^{n}_{i=1} \frac{\left( \bx_{i}-\bmu \right)\left(\bx_{i}-\bmu \right)^{T}}{\left(\bx_{i}-\bmu \right)^{T}\{\bhT(\bmu)\}^{-1}
\left(\bx_{i}-\bmu\right)} = \frac{d}{n} \sum^{n}_{i=1} \frac{\btheta_i \btheta_i^{T}}{\btheta_i^{T}\{\bhT(\bmu)\}^{-1}\btheta_i},
\end{equation}
where $\btheta_i$ is defined as in (\ref{sccm}).  In the aforementioned paper, a solution $\bhT(\bmu)$ to (\ref{tyler}) is shown to exist under general conditions and
to be readily computed via an IRLS algorithm. The solution is also unique up to a scalar multiple, and can be made unique by 
demanding for example that $\trace\{\bhT(\bmu)\} =1$ or $\det\{\bhT(\bmu)\} =1$.  Under general random sampling, $\bhT(\bmu)$ is known to be asymptotically normal, and 
when replacing $\bmu$ by a consistent estimate $\widehat{\bmu}_n$, $\bhT_n = \bhT(\widehat{\bmu}_n)$ is asymptotically equivalent to $\bhT(\bmu)$ under the same conditions 
as in Lemma \ref{lemma-asym}.   If $\widehat{\bmu}_n$ is affine equivariant, then so is $\bhT_n$ in the sense that an affine transformation of the data 
$\bx_i \rightarrow \bB \bx_i + \bb$ for $i = 1, \ldots, n$ induces the transformation $\{\beta ~\bhT_n ~|~ \beta > 0\} \rightarrow \{\beta ~\bB \bhT_n \bB^{T} ~|~ \beta > 0\}$.  

Under random sampling from an $\mE(\bmu,\bGamma;G)$ distribution, as with the \emph{SSCM}, the finite sample distribution of $\bhT(\mu)$, and hence its asymptotic
distribution, does not depend on the radial component $G$ and depends on $\bGamma$ only up to a scalar multiple. This follows since $\bhT(\mu)$ is a function of the 
random sample $\btheta_1, \ldots, \btheta_n$ which comes from an $ACG_d(\bGamma)$ distribution.  It is worth noting that Tyler's matrix corresponds to the maximum 
likelihood estimate for $\bGamma$ under random sampling from the angular central Gaussian distribution, see \cite{Tyler87b}. Another important property of
Tyler's matrix is that is minimizes the maximum asymptotic variance over all elliptical distributions for estimates of the shape component of $\bGamma$. The asymptotic 
distribution of a shape component of $\bhT_n$ under random sampling from an elliptical distribution is given by (\ref{shape}) with $\sigma_{1,\bhT_n,G} = 1+2/d$, which 
does not depend on $G$. Whereas, for any affine equivariant estimate of $\bGamma$, though, $\sup_G \sigma_{1,\widehat{\bGamma},G} \ge 1+2/d$.

\section{Theoretical Results} \label{Theory}
\subsection{The inadmissibility of the \emph{SSCM} under elliptical distributions} \label{Optimal}
Hereafter, it is presumed that $\bx_1, \ldots, \bx_n$ is a random sample from an $\mE(\bmu,\bGamma;G)$ distribution. As noted in section \ref{subSSCM}, the
\emph{SSCM} is a consistent estimate of $\bXi$, but $\bXi$ itself is not a multiple of $\bGamma$. However, $\bhS_n$ can be transformed into a consistent
estimate of a multiple of $\bGamma$ in the following manner. Consider the spectral value decomposition for $\bGamma = \bQ\bLambda\bQ^T$, where $\bQ$ is an 
orthogonal matrix whose columns consists of an orthonormal set of eigenvectors of $\bGamma$, and  $\bLambda$ is a diagonal matrix whose diagonal elements
$\lambda_1 \ge \ldots \ge \lambda_d$ are the ordered eigenvalues of $\bGamma$. The spectral value decomposition of the matrix $\bXi$, on the other hand, 
is known to have the form $\bXi = \bQ\bDelta\bQ^T$ where $\bQ$ is the same as that for $\bGamma$ and $\bDelta = diagonal\{\phi_1, \ldots, \phi_d\}$  with
$\phi_1 \ge \ldots \ge \phi_d$ being the ordered eigenvalues of $\bXi$. The relationship between $\bDelta$ and $\bLambda$ is given by
\begin{equation} \label{evalues}
\phi_j = E\left[ \frac{\lambda_j ~\chi^2_{1,j}}{\sum_{r=1}^d \lambda_r ~\chi^2_{1,r}}\right], \quad \mbox{for} \quad j = 1, \dots, d,
\end{equation}
with $\chi^2_{1,1}, \ldots, \chi^2_{1,d}$ being mutually independent chi-square distributions on $1$ degrees of freedom,
see e.g. \cite{Boente99} or \cite{Taskinen12}. Hence the eigenvalues of the \emph{SSCM}, which consistently estimate the eigenvalues of $\bXi$, 
are not consistent estimates of the eigenvalues of $\bGamma$. However, (\ref{evalues}) implies $\bDelta$ is a function of $\bLambda$, which we
denote by $\bDelta = h(\bLambda)$.  Without loss of generality, since $\bGamma$ is confounded with the radial component $G$, we presume $\bGamma$
is normalized so that $\trace(\bGamma) = 1$, and since $\trace(\bXi) =1$ the function $h$ can be viewed as a function from $\mathbb{R}^{d-1}$ to $\mathbb{R}^{d-1}$.
The function $h$ can be shown to be one-to-one and continuously differentiable. This then implies that $h(\bGamma) = \bQ h(\bLambda)\bQ^T$ is one-one and
continuously differentiable, and hence $h^{-1}(\bhS_n)$ is a consistent and asymptotically normal estimate of $\bGamma$, i.e.\
\begin{equation} \label{normS}
\sqrt{n} ~ vec\{ h^{-1}(\bhS_n) - \bGamma \} \sim_a \sqrt{n} ~ vec\{ h^{-1}(\bhS(\bmu)) - \bGamma \} \rightarrow_\mD Normal_{d^2}\left(\bzero, \mV_{h^{-1}(\mS)}(\bGamma) \right).
\end{equation}
If we also normalize Tyler's scatter matrix so that $\trace(\bhT_n) =1$ we then also have
\begin{equation} \label{normT}
\sqrt{n} ~ vec\{ \bhT_n - \bGamma \} \sim_a \sqrt{n} ~ vec\{ \bhT(\bmu) - \bGamma \} \rightarrow_\mD Normal_{d^2}\left(\bzero, \mV_\mT(\bGamma) \right).
\end{equation}
The forms of asymptotic variances are discussed later. From the general theory of maximum likelihood estimation, though, it can be shown that
the \emph{SSCM} is asymptotically inadmissible in the sense stated in the following theorem. Here, for two symmetric matrices of order $d$, the notation
$\bA_1 > \bA_2$ and $\bA_1 \ge \bA_2$ implies $\bA_1 - \bA_2$ is positive definite and positive semi-definite respectively, which is the usual partial
ordering for symmetric matrices.

\begin{theorem}  \label{inadmiss}
Let $\bx_{1}, \ldots ,\bx_{n}$ represent an i.i.d.\ sample from a $\mE(\bmu,\bGamma;G)$ distribution. Then for $\mV_{h^{-1}(\mS)}(\bGamma)$
and  $\mV_{\mT}(\bGamma)$ defined in (\ref{normS}) and (\ref{normT}) respectively, we have $\mV_{h^{-1}(\mS)}(\bGamma) \ge \mV_{\mT}(\bGamma)$ for all
$\bGamma > \bzero$ and $\mV_{h^{-1}(\mS)}(\bGamma) \ne \mV_{\mT}(\bGamma)$ for some $\bGamma > \bzero$.
\end{theorem}

\emph{PROOF.}
Since $h^{-1}(\bhS_n)$ and $h^{-1}(\bhS(\bmu))$ as well as  $\bhT_n$ and $\bhT(\bmu)$ are asymptotically equivalent, it is sufficient
to compare $h^{-1}(\bhS(\bmu))$ to $\bhT(\bmu)$. As noted previously, both of these estimates are functions of $\btheta_1, \ldots \btheta_n$ which
represents an i.i.d.\ sample from an $ACG_d(\bGamma)$ distribution. As shown in \cite{Tyler87b}, though, the maximum likelihood estmate of $\bGamma$ is
$\bhT(\bmu)$. The $ACG_d(\bGamma)$ distribution also satisfies sufficient regularity conditions to establish that its maximum likelihood estimate is
asymptotically efficient, i.e.\ that its asymptotic variance-covariance matrix equals the inverse of its Fisher information matrix.  Furthermore,
it can be shown that the joint limiting distributions in (\ref{normS}) and (\ref{normT}) are jointly multivariate normal. Consequently, the theorem
follows from the general theory of maximun likelihood estimate, see e.g.\ Theorem 4.8 in \cite{Lehmann98}.  $\Box$

\begin{remark}
Consider the spherical case, i.e.\ $\bGamma \propto \bI_d$. \mbox{For this case,  $\bXi = E[\btheta \btheta^T] = d^{-1} \bI_d$} and 
$E[\btheta \btheta^T \otimes \btheta \btheta^T] = \{d(d+2)\}^{-1}\{\bI_{d^2} + \mK_{d,d} + vec(\bI_d)vec(\bI_d)^T\}$, see e.g.\ \cite{Tyler87a}.  
\mbox{Consequently}, the asymptotic variance of the \emph{SSCM} given in (\ref{clt}) can be expressed as  $\mV_\mS =\{d(d+2)\}^{-1} \mM $, where 
$\mM =\{\bI_{d^2} + \mK_{d,d} - (2/d) vec(\bI_d)vec(\bI_d)^T\}$. By comparison, it is shown in \cite{Tyler87a} that 
$\mV_{\mT}(d^{-1}\bI_d) = \{(d+2)/d^3\} \mM$. Although $\{d(d+2)\}^{-1} < \{(d+2)/d^3\} $, this does not contradict the asymptotic optimality
of the maximum likelihood estimate $\bhT_n$ since $\bhS_n$ itself is not consistent for $\bGamma/\trace(\bGamma)$ when $\bGamma \not\propto \bI_d.$  
More generally, for an asymptotically normal affine equivariant estimate of scatter, say $\widehat{\bGamma}$, under a spherical distribution 
$\sqrt{n} ~vec\{\widehat{\bGamma}/\trace(\widehat{\bGamma}) -d^{-1}\bI_d\} \rightarrow_\mD
Normal_{d^2}\left(\bzero, d^{-2}\sigma_{1,\widehat{\bGamma},G}\mM \right)$ where $\sigma_{1,\widehat{\bGamma},G}$ is the same as in (\ref{shape}).
For the sample variance-covariance matrix, $\sigma_{1,\bS_n,G} = 1$ at the multivariate normal distribution and since
$1/d^{2} > \{d(d+2)\}^{-1}$, its asymptotic variance is also greater than that of the $\emph{SSCM}$. 
Essentially, the \emph{SSCM} itself can be viewed as a super efficient estimate of shape when $\bGamma \propto \bI_d$ but at a cost of inconsistency
otherwise.
\end{remark}

\subsection{Asymptotic Calculations} \label{Hyper}
Theorem \ref{inadmiss} of the last section states that the asymptotic variance of a consistency adjusted \emph{SSCM} is greater than that
of the \emph{ASSCM}, i.e.\ of Tyler's scatter matrix. In this section, the degree of the asymptotic inefficiency of the \emph{SSCM} relative
to the \emph{ASSCM} is studied. As noted previously, although the \emph{SSCM} is an inconsistent estimator of $\bGamma$, its eigenvectors 
are consistent estimates of the eigenvectors of $\bGamma$. Because of this result, along with the orthogonal equivariance of the \emph{SCCM}, the 
eigenvectors of the \emph{SSCM} have been recommended as robust estimates of the principal component directions, most notably in \cite{Locantore99}, 
\cite{Marden99} and \cite{Maronna06}. Consequently, we focus our study here on the degree of inefficiency of the eigenvectors of the \emph{SSCM}. Ironically, 
PCA is primarily of interest when the eigenvalues of $\bGamma$ are well separated, but as shown here, this is the case when the \emph{SSCM} is least
efficient. 

Before stating our results formally, additional notation is needed. Recall the spectral value decompositions of $\bGamma = \bQ\bLambda\bQ^T$ and $\bXi = \bQ\bDelta\bQ^T$,
with $\lambda_1 \ge \ldots \ge \lambda_d$ and $\phi_1 \ge \ldots \ge \phi_d$ being the eigenvalues of $\bGamma$ and $\bXi$ respectively and the columns of $\bQ$ being
the corresponding eigenvectors of both $\bGamma$ and $\bXi$.  The asymptotic distribution of the estimated eigenvectors depends on the multiplicities of the eigenvalues, 
and so denote the distinct eigenvalues of $\bGamma$ as $\lambda_{(1)} > \ldots > \lambda_{(m)}$ and of $\bXi$ as $\phi_{(1)} > \ldots > \phi_{(m)}$, with the respective 
multiplicities being $d_1, \ldots, d_m$. Note that equation (\ref{evalues}) implies that both the order and the multiplicities of the eigenvalues of $\bGamma$ and $\bXi$ are the same.
The matrix of eigenvectors $\bQ = [ \bq_1, \cdots, \bq_d]$ is not uniquely defined, especially when multiple roots exists. However the subspace spanned by the eigenvectors associate
with $\lambda_{(j)}$, and the corresponding \emph{eigenprojection} onto the subspace, is uniquely defined. That is, the spectral value decompositions have the unique
representations $\bGamma = \sum_{j=1}^m \lambda_{(j)} \bP_{j}$ and $\bXi = \sum_{j=1}^m \phi_{(j)} \bP_{j}$, where $\bP_{j} = \sum_{k = m_j +1}^{ m_j + d_j}\bq_k\bq_k^T$ and
$m_1 = 0$ and $m_j = d_1 + \ldots + d_{j-1}$ for $j = 2, \ldots, m$. Consequently, it is simpler to work with the asymptotic distributions of the eigenprojections rather than
the asymptotic distribution of the eigenvectors themselves.

Let the spectral value decomposition for $\bhT_n$ and $\bhS_n$ be represented respectively by \mbox{$\bhT_n = \widehat{\bQ}_\mT\widehat{\bLambda}\widehat{\bQ}_\mT^T$ }
and $\bhS_n = \widehat{\bQ}_\mS\widehat{\bDelta}\widehat{\bQ}_\mS^T$.  For multivariate distributions absolutely continuous with respect to Lebesgue measure in $\mathbb{R}^d$, 
the eigenvalues of $\bhT_n$ and of $\bhS_n$ are distinct with probability 1. For $k = 1, \ldots, d$, let $\widehat{\bq}_{\mT,k}$ and 
$\widehat{\bq}_{\mS,k}$ represent the corresponding normalized eigenvectors, i.e.\ the columns of $\widehat{\bQ}_\mT$ and  $\widehat{\bQ}_\mS$ respectively.  
Consistent estimates for the eigenprojections $\bP_{j}$ are then given by
$\bhP_{\mT,j} = \sum_{k = m_j +1}^{ m_j + d_j}\widehat{\bq}_{\mT,k}\widehat{\bq}_{\mT,k}^T$ and
$\bhP_{\mS,j} = \sum_{k = m_j +1}^{ m_j + d_j}\widehat{\bq}_{\mS,k}\widehat{\bq}_{\mS,k}^T$ for $j = 1, \ldots, m$. Consistency follows since
eigenprojections are continuous functions of their matrix argument. More generally, eigenprojections are also analytic functions of their matrix argument, see e.g.\ \cite{Kato66}. 
Thus, the asymptotic distribution for the eigenprojections can then be obtained using the delta method. This gives
$\sqrt{n} ~ vec\{ \bhP_{\mT,j} - \bP_{j} \}  \rightarrow_\mD Normal_{d^2}\left(\bzero, \mV_{\bP_{\mT,j}}(\bGamma) \right)$ and 
$\sqrt{n} ~ vec\{ \bhP_{\mS,j} - \bP_{j} \}  \rightarrow_\mD Normal_{d^2}\left(\bzero, \mV_{\bP_{\mS,j}}(\bGamma) \right)$, for $j = 1, \ldots, m$. The asymptotic variance 
covariance matrices, derived in Appendix \ref{asy-dis}, are respectively
\begin{equation} \label{varP}
\mV_{\bP_{\mT,j}}(\bGamma) =  \sum_{k =1,k \ne j}^m \alpha_{\mT,j,k}\mM_{j,k} \quad \mbox{and} \quad \mV_{\bP_{\mS,j}}(\bGamma) =  \sum_{k =1,k \ne j}^m \alpha_{\mS,j,k}\mM_{j,k},
\end{equation}
where $\mM_{j,k} = (1/2)(I + \mK_{d,d}) \left(\bP_j \otimes \bP_k + \bP_k \otimes \bP_j \right)$, 
\begin{equation} \label{alpha}
\alpha_{\mT,j,k} =  \frac{d+2}{d} \frac{2\lambda_{(j)}\lambda_{(k)}}{(\lambda_{(j)} - \lambda_{(k)})^2} \quad \mbox{and} \quad
\alpha_{\mS,j,k} = \frac{2\psi_{(j,k)}}{(\phi_{(j)} - \phi_{(k)})^2}~.
\end{equation}
The terms
\begin{equation} \label{psi}
\psi_{(j,k)} = \frac{1}{d_j d_k}E\left[ \frac{\lambda_{(j)}\lambda_{(k)} ~\chi^2_{(j)}\chi^2_{(k)}}{\left\{\sum_{r=1}^m \lambda_{(r)} ~\chi^2_{(r)}\right\}^2}\right] \quad \mbox{and}
\quad \phi_{(j)} = \frac{1}{d_j} E\left[ \frac{\lambda_{(j)} ~\chi^2_{(j)}}{\sum_{r=1}^m \lambda_{(r)} ~\chi^2_{(r)}}\right],
\end{equation}
where $\chi^2_{(r)}, r= 1, \ldots, m$ have mutually independent chi-squares distributions on $d_r$ degrees of freedom respectively.  The form of the eigenvalue $\phi_{(j)}$ given above
is equivalent to equation (\ref{evalues}).

The forms of the asymptotic variance-covariance matrices given in (\ref{varP}) correspond to their spectral value decompositions since the matrices $\mM_{j,k}$ are 
symmetric idempotent matrices, i.e.\ they are the eigenprojections associated with both $\mV_{\bP_{\mT,j}}(\bGamma)$ and $\mV_{\bP_{\mS,j}}(\bGamma)$.
The corresponding eigenvalues are respectively $\alpha_{\mT,j,k}$ and  $\alpha_{\mS,j,k}$, with multiplicities $d_j d_k$, for $k \ne j, k = 1, \ldots, m$. The rank of the $d^2 \times d^2$ 
matrices $\mV_{\bP_{\mT,j}}(\bGamma)$ and $\mV_{\bP_{\mS,j}}(\bGamma)$ are thus $d_j(d-d_j)$. The results of the previous section, in particular Theorem \ref{inadmiss}, imply that
$\mV_{\bP_{\mT,j}}(\bGamma) \le \mV_{\bP_{\mS,j}}(\bGamma)$ and consequently $\alpha_{\mT,j,k} \le \alpha_{\mS,j,k}$. We believe the last inequality to always be 
strict. It appears though to be quit difficult to derive this inequality directly from the expressions given in (\ref{alpha}).

To help obtain insight into how the asymptotic efficiency of the \emph{SSCM} is affected whenever the underlying elliptical distribution is not spherical, we hereafter
consider the special case in which $\bGamma$ has only two distinct eigenvalues. That is, suppose $m=2$ with $\lambda_{(1)}$ and $\lambda_{(2)}$ having multiplicities $d_1$ and 
$d_2 = d-d_1$ respectively.  The asymptotic efficiencies depend on the values of $\lambda_{(1)}$ and $\lambda_{(2)}$ only through its ratio $\rho^2 = \lambda_{(2)}/\lambda_{(1)}$.  
Also, it is sufficient to consider only the eigenprojection $\bP_1$ since $\bP_2 = \bI_d -\bP_1$. The asymptotic variance-covariance matrices reduce to
$\mV_{\bP_{\mT,1}}(\bGamma) = \alpha_{\mT}(\rho)\mM_{1,2}$ and $\mV_{\bP_{\mS,1}}(\bGamma) = \alpha_{\mS}(\rho)\mM_{1,2}$, where $\alpha_{\mT}(\rho) = \alpha_{\mT,1,2}$ and 
$\alpha_{\mS}(\rho) = \alpha_{\mS,1,2}$. Hence $\mV_{\bP_{\mT,1}}(\bGamma) \propto \mV_{\bP_{\mS,1}}(\bGamma)$ and so the asymptotic efficiency of $\bhP_{\mS,1}$ relative to 
$\bhP_{\mT,1}$, or equivalently the the asymptotic efficiency of $\bhP_{\mS,2}$ relative to  $\bhP_{\mT,2}$, can be reduced to the scalar value
\begin{equation} \label{are}
ARE_{d,d_1}\left(\bhP_{\mS,1},\bhP_{\mT,1};\rho\right) = \frac{\alpha_{\mT}(\rho)}{\alpha_{\mS}(\rho)}.
\end{equation} 
For this case, expressions for $\psi_{(j,k)}$ and $\phi_{(j)}$ in terms of the Gauss-hypergeometric functions $_2F_1$ 
are given by
\begin{equation} \label{hyp-psi}
\psi_{(1,2)} = \psi_{(2,1)} =  {}_{2}F_{1}\left(2, \frac{d_{2}+2}{2}; \frac{d+4}{2}; 1-\rho^2 \right) \frac{\rho^2}{d\left(d+2\right)},
\end{equation}
\begin{equation} \label{hyp-phi}
\phi_{(1)} = {}_{2}F_{1}\left(1, \frac{d_{2}}{2}; \frac{d+2}{2}; 1-\rho^2 \right)\frac{1}{d}\quad \mbox{and} \quad
\phi_{(2)} = {}_{2}F_{1}\left(1, \frac{d_{2}+2}{2}; \frac{d+2}{2}; 1-\rho^2 \right) \frac{\rho^2}{d}. 
\end{equation}
Using properties of the hypergeometric functions then yields
\begin{equation} \label{hyp-are}
ARE_{d,d_1}\left(\bhP_{\mS,1},\bhP_{\mT,1};\rho\right) = \frac{{}_{2}F^{2}_{1}\left(1, \frac{d_{2}+2}{2}; \frac{d+4}{2}; 1-\rho^2\right)}
{{}_{2}F_{1}\left(2, \frac{d_{2}+2}{2}; \frac{d+4}{2}; 1-\rho^2\right)}.
\end{equation} 
The derivations of (\ref{hyp-psi}), (\ref{hyp-phi}) and  (\ref{hyp-are}) are given in 
section \ref{app-hyper} of the appendix.

The formulas given in (\ref{hyp-psi}), (\ref{hyp-phi}), and (\ref{hyp-are}) simplify in the two dimensional case, i.e.\ for $d =2, d_1 =1$ and $d_2 = 1$. 
For this case, it is shown in section \ref{app-hyper} of the appendix that
\begin{equation} \label{dim2}
\phi_{(1)} = \frac{1}{1+\rho}, \quad \phi_{(2)} = \frac{\rho}{1+\rho},  \quad \mbox{and} \quad \psi_{(1,2)} = \frac{\rho}{2(1+\rho)^2}.
\end{equation}
This then gives $\alpha_\mS(\rho) = \rho/(1-\rho)^2$, and since $\alpha_\mT(\rho) = 4\rho^2/(1-\rho^2)^2$ we have
$ARE_{2,1}\left(\bhP_{\mS,1},\bhP_{\mT,1};\rho\right) = 4\rho/(1+\rho)^2$, which is displayed in Figure 1. Note that the
the asymptotic relative efficiency goes to $0$ as $\rho$ approaches $0$.  Small values of $\rho$, though, correspond to situations when
one is most interested in principal components analysis. 
\begin{figure}[h] \label{are2}
  \centering
   \includegraphics[scale= 0.25]{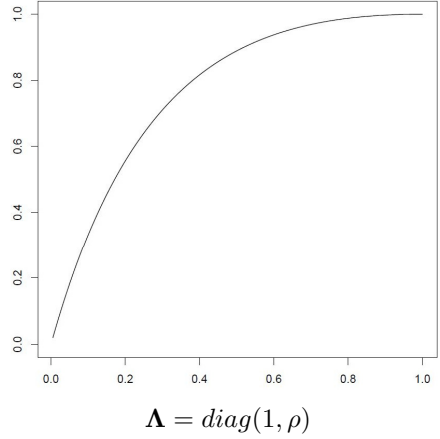} \\ $\bLambda = diag(1,\rho)$
    \caption{Asymptotic efficiency in $\mathbb{R}^{2}$, as a function of $\rho$, of the eigenprojection of \emph{SSCM} relative to that of \emph{ASSCM}.}
\end{figure}

In three dimensions, $d=3$,  the special case we are considering has two sub-cases, namely  $\lambda_{2} = \lambda_{3}$ and $\lambda_{1} = \lambda_{2}$, i.e.\ $(d_{1},d_2) = (1,2)$ and 
$(2,1)$ respectively.  Presented in Figure 2 are the asymptotic relative efficiencies (\ref{hyp-are}) under both scatter structures as a function of $\rho$.  As in two dimensions, 
the asymptotic relative efficiency is low when $\rho$ is close to 0.  Of the two scatter structures considered, the case $(d_1,d_2) = (1,2)$ is less favorable for the \emph{SSCM}. 

\begin{figure}[h] \label{are3}
\begin{tabular}{cc} 
\includegraphics[scale= 0.25]{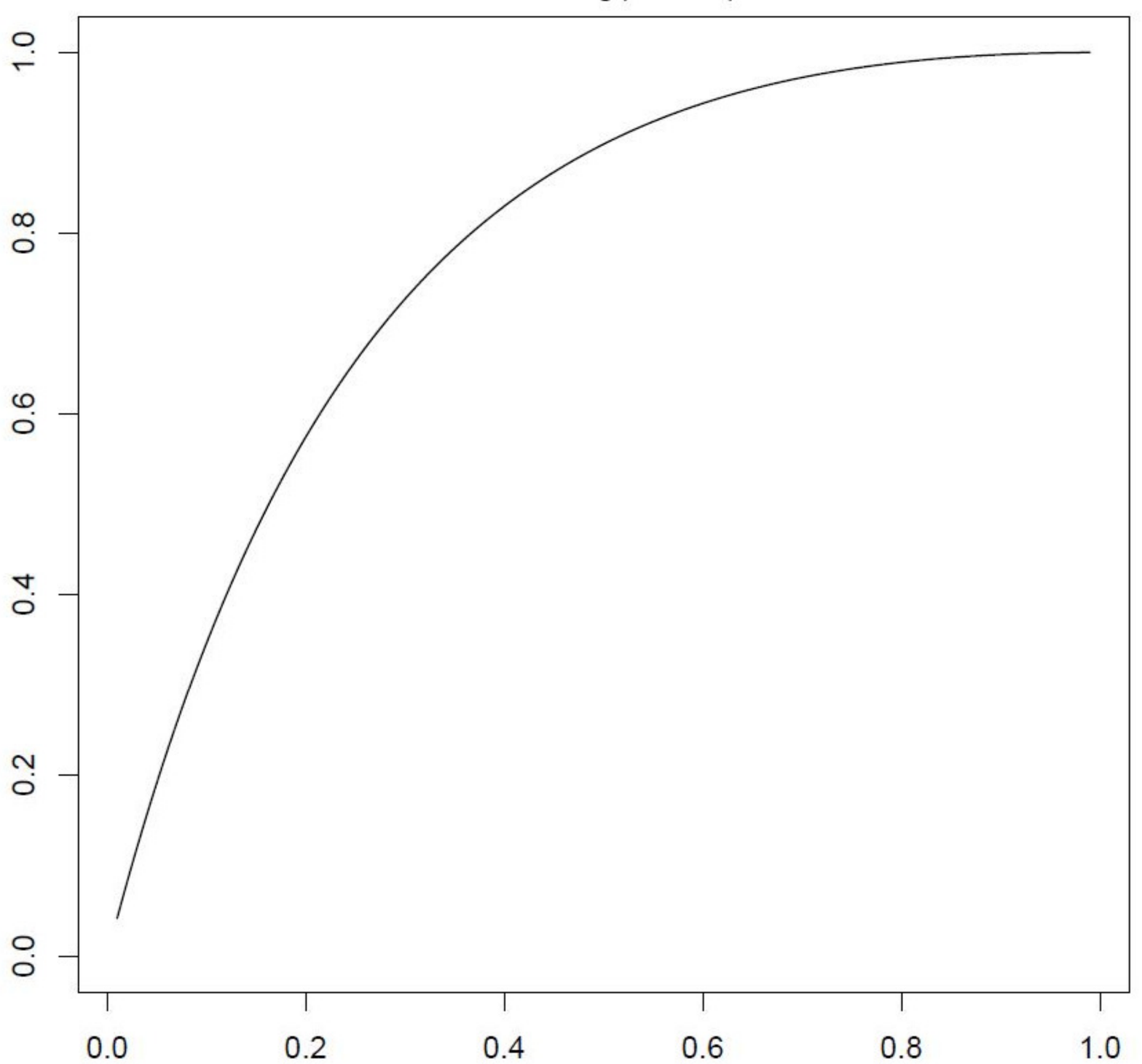} &
\includegraphics[scale= 0.25]{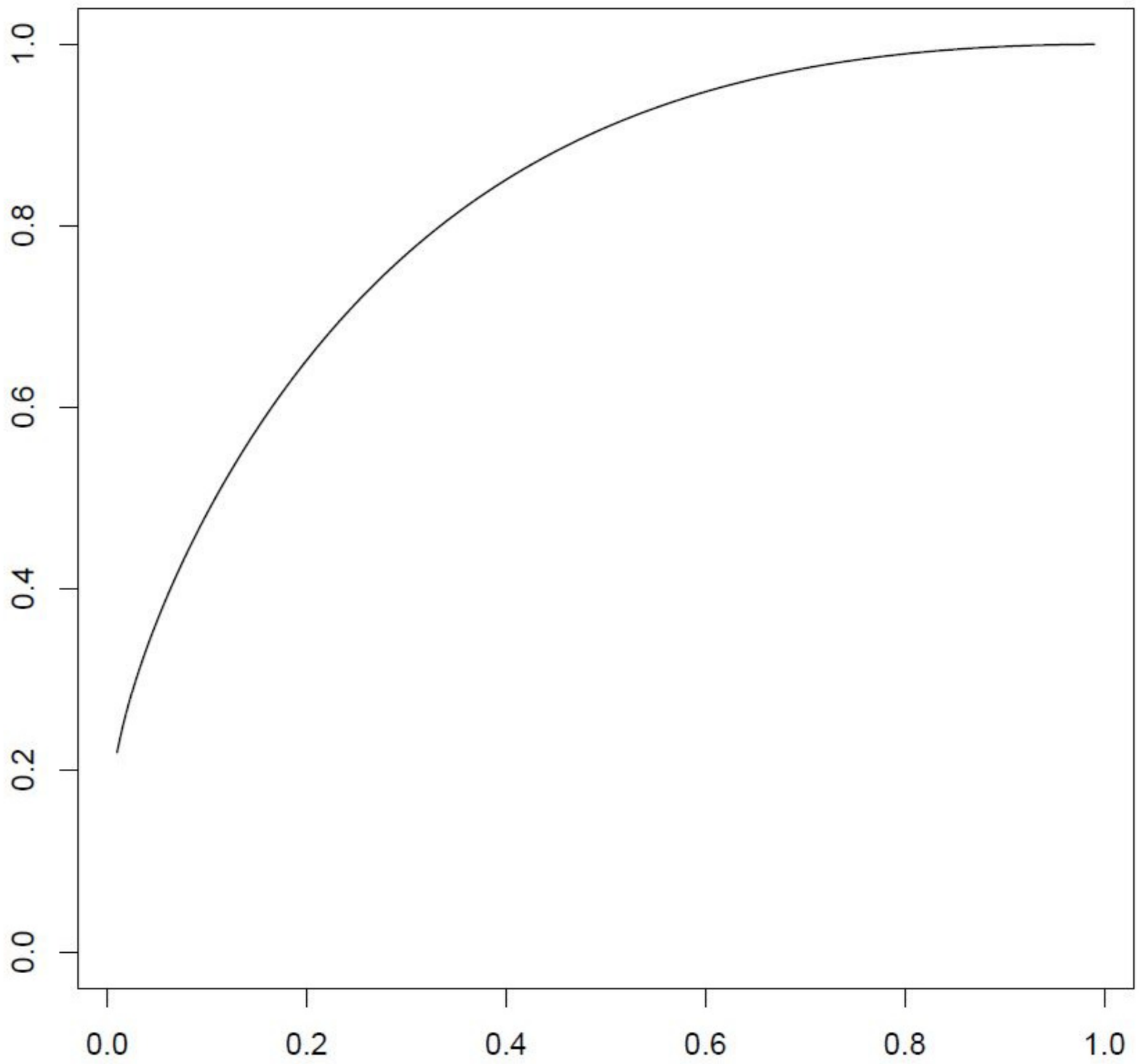} \\
$\bLambda = diag(1,\rho,\rho)$ &  $\bLambda = diag(1,1,\rho)$
\end{tabular}
    \caption{Asymptotic efficiency in $\mathbb{R}^{3}$, as a function of $\rho$, of the eigenprojection of \emph{SSCM} relative to that of \emph{ASSCM}. }
\end{figure}

Plotted in Figure 3 are the asymptotic relative efficiencies when the dimension is $d = 5$. The sub-cases here are now $(d_{1},d_2) = (1,4), (2,3), (3,2)$ and $(4,1)$. From the
plots it can be noted that the larger the dimension of the principal component space associated with the larger eigenvalue, the higher the asymptotic relative efficiency,
with the asymptotic relative efficiency in the last sub-case not being too low even when $\rho$ approaches zero. This indicates not much is lost by using the \emph{SSCM} for this case,
even under a nearly singular scatter structure.  However, as the dimension of the principal component space decreases, the asymptotic relative efficiencies steadily decreases, 
and can be quite poor for nearly singular scatter structures.   

\begin{figure}[h] \label{are5}
\begin{tabular}{cc} 
\includegraphics[scale= 0.25]{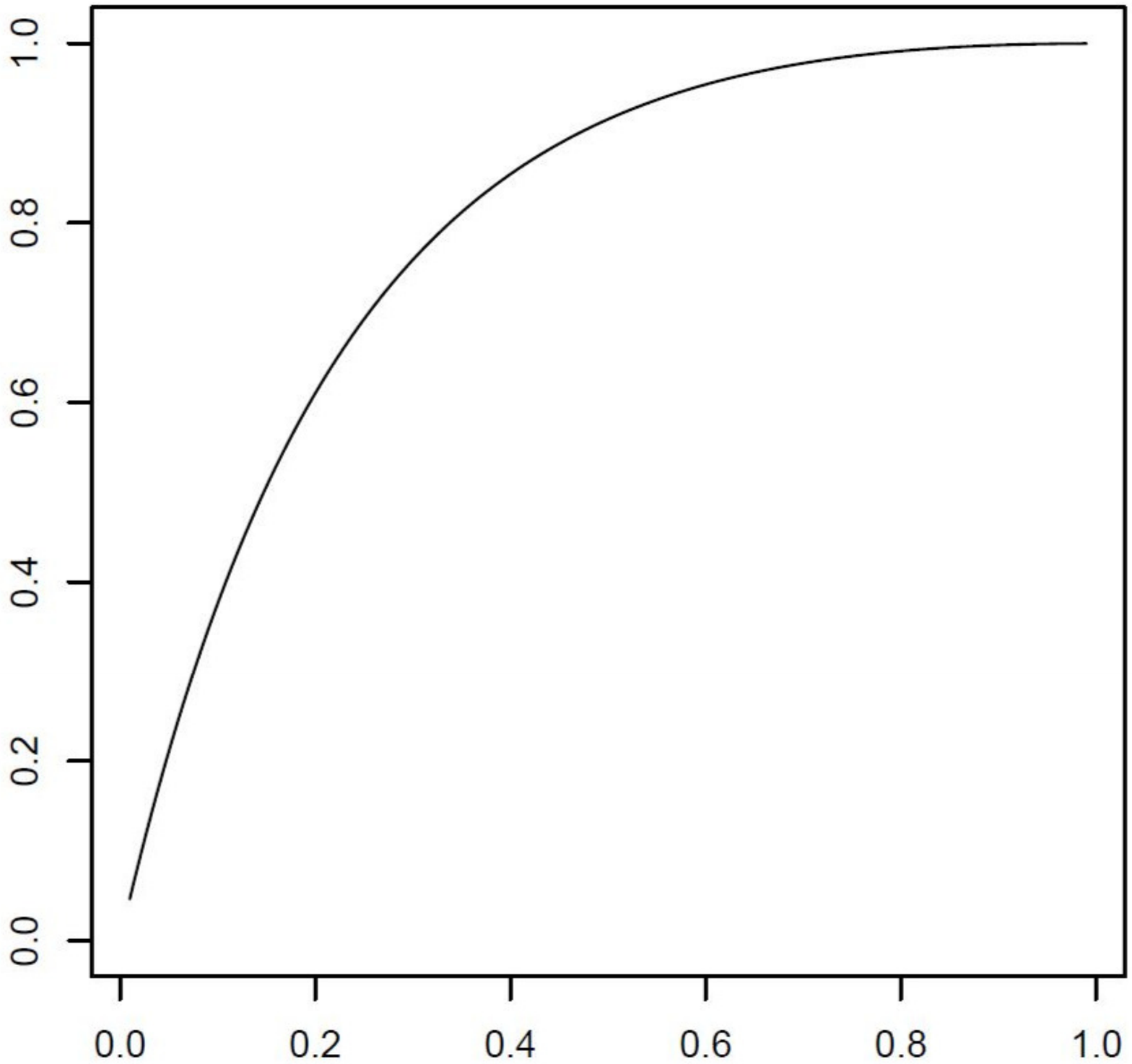} &
\includegraphics[scale= 0.25]{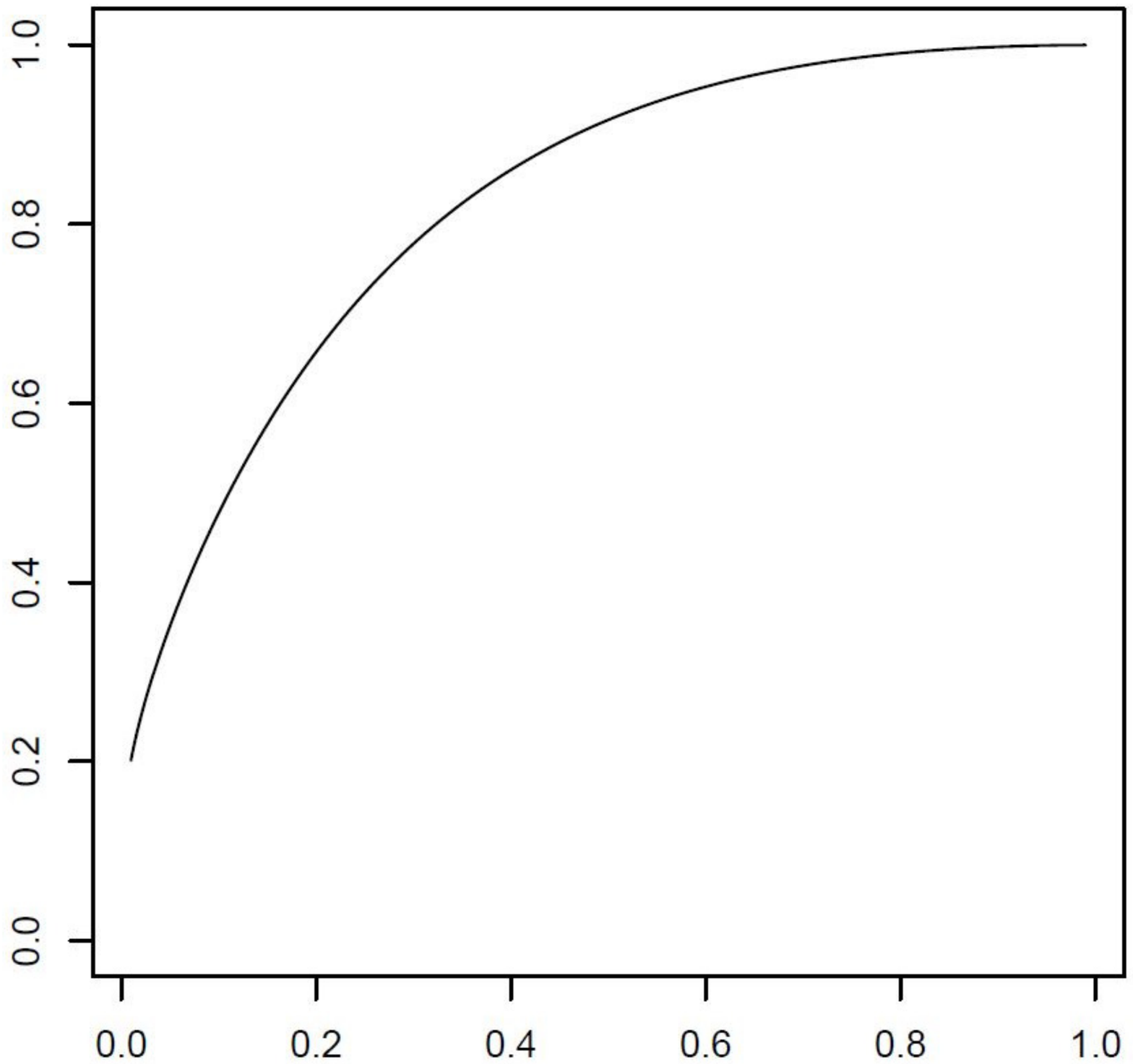} \\
$\bLambda = diag(1,\rho,\rho,\rho,\rho)$ &  $\bLambda = diag(1,1,\rho,\rho,\rho)$ \\[1cm]
\includegraphics[scale= 0.25]{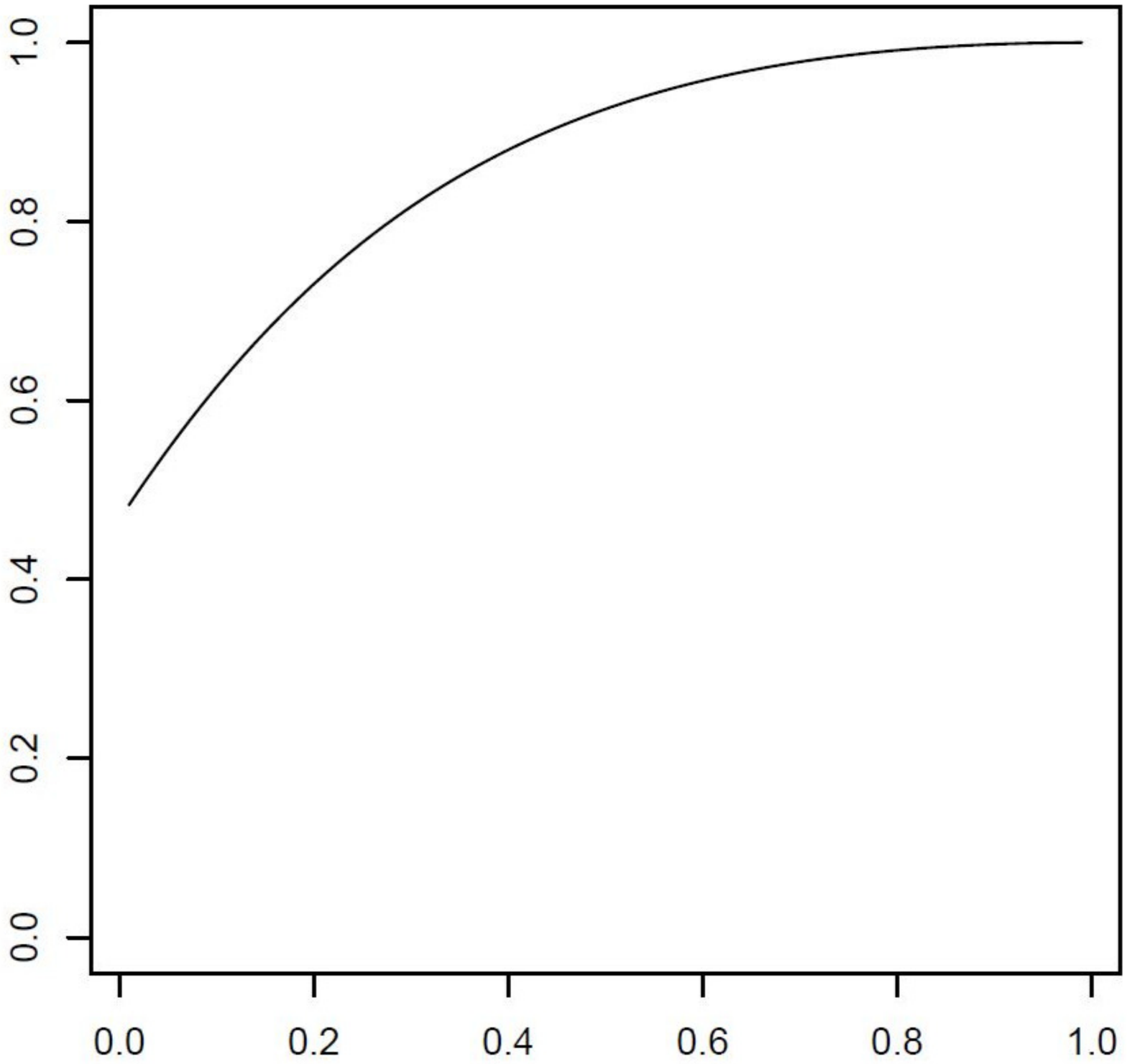} &
\includegraphics[scale= 0.25]{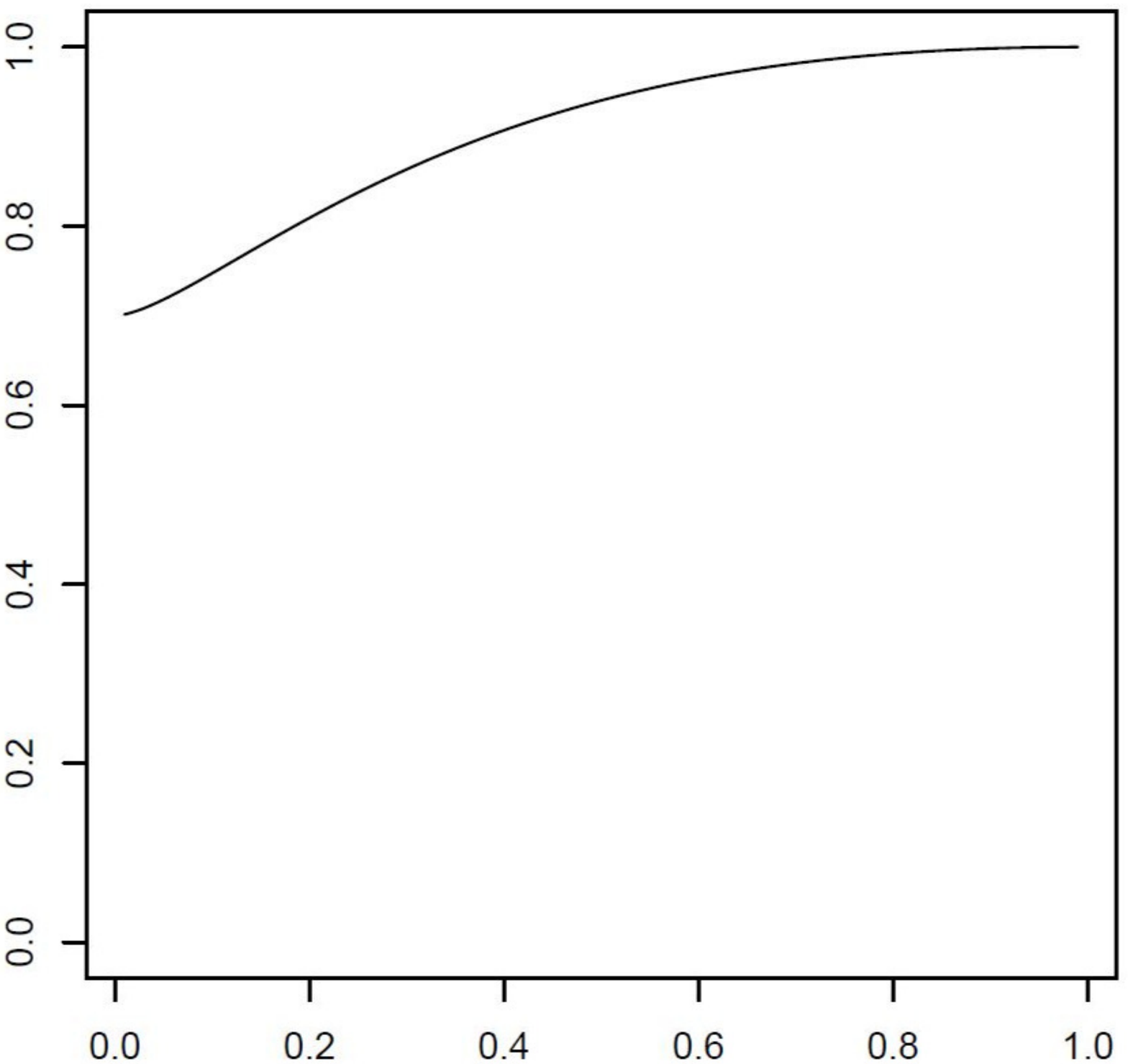} \\
$\bLambda = diag(1,1,1,\rho,\rho)$ &  $\bLambda = diag(1,1,1,1,\rho)$
\end{tabular}
    \caption{Asymptotic efficiency in $\mathbb{R}^{5}$, as a function of $\rho$, of the eigenprojection of \emph{SSCM} relative to that of \emph{ASSCM}. }		
\end{figure}

In general, since ${}_2F_1(a,b;c;0) = 1$, the asymptotic relative efficiency (\ref{hyp-are}) goes to one as \mbox{$\rho \rightarrow 1$}. Also, as shown in Appendix \ref{app-hyper}
\begin{equation} \label{rho-0}
 \lim_{\rho  \rightarrow 0} ARE_{d,d_1}\left(\bhP_{\mS,1},\bhP_{\mT,1};\rho\right) = \left\{ 
\begin{array}{ccl} \left(1+\frac{2}{d}\right)\left(1-\frac{2}{d_1}\right)  & \quad \mbox{for} \quad & d_1 > 2 \\ 0 & \quad \mbox{for} \quad & d_1 = 1, 2. \end{array}  \right. 
\end{equation}
A small value of $\rho$ implies the principal component 
space associated with the larger root is well separated from the principal component space associated with the smaller root. For this case, in high dimensions, for the 
asymptotic relative efficiency (\ref{hyp-are}) to be greater than $0.80$ or $0.90$, the dimension of the principal component space needs to be at least approximately 
$d_1 \ge 10$ and $d_2 \ge 20$ respectively. Consequently, the \emph{SSCM} is not particular efficient in cases where principal components analysis 
is used to reduce the dimensionality of the data to less than ten.

For the scatter structures considered here, to find the asymptotic relative efficiency of the \emph{SSCM} to an affine equivariant estimate of scatter $\widehat{\bGamma}$ other 
than Tyler's scatter matrix, one simply needs to multiply (\ref{hyp-are}) by the asymptotic efficiency of Tyler's scatter matrix relative to $\widehat{\bGamma}$, which gives
 $\sigma_{1,\widehat{\bGamma},G} ~ d/(d+2)$. The term $\sigma_{1,\widehat{\bGamma},G}$ is defined as in (\ref{shape}). Note that in contrast to (\ref{hyp-are}), the
term $\sigma_{1,\widehat{\bGamma},G}$ does not depend on the values of $\bGamma$ but does depend on the underlying elliptical distribution.  Expressions for the 
$\sigma_{1,\widehat{\bGamma},G}$  for the \emph{M}-estimates of scatter are given in \citep{Tyler83}, and for the high breakdown point
\emph{S}-estimates and re-weighted \emph{S}-estimates by \citep{Lopuhaa89,Lopuhaa99}.  The expressions given for the \emph{M}-estimates also apply to the
high breakdown point \emph{MM}-estimates. For the sample covariance matrix,  $\sigma_{1,\bS_n,G} = (1+2/d) E[R_G^4]/E[R_G^2]^2$, which is one under multivariate
normality.

\section{Finite Sample Performance} \label{Sim}
The results of the previous section showed that asymptotically, the inefficiency of the \emph{SSCM} relative to the \emph{ASSCM} can be quite severe under an elliptical model 
which is far from spherical. In this section, the finite sample efficiencies are considered. In the finite sample setting, the behavior of both the \emph{SSCM} and the
\emph{ASSCM} depend upon the choice of the estimate of location.  Also, when estimating location, the finite sample distributions are dependent on the particular elliptical
family $\mE(\bmu,\bGamma;G)$, i.e.\ upon $G$. Here we focus on the estimate based on known $\bmu$.  For this case, the finite sample distributions of $\bhS(\bmu)$ and of
$\bhT(\bmu)$ do not depend upon $G$. Consequently, for simulation purposes, we only then need to consider random sampling from multivariate normal distributions, even if the true underlying
elliptical distribution does not possess any moments.  While it is hardly ever the case that the location vector is known a priori, considering this case allows us to focus
on the effect of the covariance or scatter structure on the finite sample properties of the estimates.

As in section \ref{Hyper}, we consider the eigenprojections of the estimates only and do so under the simplified (yet informative) case for which there are just two distinct
eigenvalues of multiplicity $d_1$ and $d_2 = d-d_1$ respectively. Recall that for this case, the asymptotic variances of the eigenprojections of $\bhS(\bmu)$ and $\bhT(\bmu)$ under
an elliptical distribution are proportional to each other and so the asymptotic comparison of the two estimates is reduced to a single number given by (\ref{are}).  The finite sample 
variance-covariance matrices of these eigenprojection estimators, however, do not possess the same form as their asymptotic counterparts and are not necessarily proportional to each 
other.  A natural way to compare the eigenprojection estimators to the theoretical eigenprojects is to use the concept of principal (canonical) angles between subspaces.

The notation used in the proceeding paragraphs is consistent with that utilized in \cite {Miao92}.  Principal angles can be used to describe how far apart one linear subspace is from another.  Let $L$ and $M$ be linear subspaces of $\mathbb{R}^{d}$ with $dim\left(L\right) = l \leq dim\left(M\right) = m$.  The principal angles between $L$ and $M$,
$
0 \leq \vartheta_{1} \leq \vartheta_{2} \leq \cdots \leq \vartheta_{l} \leq \pi/2
$
are given by
\begin{equation} \label{pangle} \hspace*{.7cm}
cos\: \vartheta_{i} = \frac{\left\langle {\bx}_{i}, \by_{i}\right\rangle}{\|\bx_{i}\| ~ \|\by_{i}\|} = 
\max\left\{ \frac{\left\langle \bx, \by\right\rangle}{\left\|\bx\right\| \left\|\by\right\|}: \  \bx \in L, \  \bx \perp \bx_{k}, \  \by \in M, \  \by \perp \textbf{y}_{k}, \  k = 1, \ldots, i-1\right\},
\end{equation}        
see e.g.\ \cite {Afriat57}.  If $l = m = 1$, then the sole principal angle is simply the smallest angle between two lines. It follows from the above definition that when the two subspaces coincide (i.e.\ $L = M$) then the principal angles are all $0$. In general, the number of non-zero principal angles is at most $\min(l,d-l)$ with the non-zero princpal angles between
$L$ and $M$ being the same as those between $L^\perp$ and $M^\perp$. The principal angles between two linear subspaces are
the maximal invariants under orthogonal transformations of the subspaces. That is, a function $D(L,M) = D(\bQ L,\bQ M)$ for any orthogonal transformation $\bQ$, where  
$\bQ L = \{\bx = \bQ\by : \by \in L\}$, if and only if it is a function of the principal angles between $L$ and $M$.  The following lemma presents a result given in \cite {Bjork73} 
which is useful for computing the principal angles between two linear subspaces.
\begin{lemma} \label{svd}
Let the columns of $Q_{L} \in \mathbb{R}^{n \times l}$ and $Q_{M} \in \mathbb{R}^{n \times m}$ be orthonormal bases for $L$ and $M$ respectively, and let
$\sigma_{1} \geq \sigma_{2} \geq \cdots \geq \sigma_{l} \geq 0$
be the singular values of $Q^{T}_{M}Q_{L}$, then
$cos\: \vartheta_{i} = \sigma_{i}$ for $i = 1, \ldots, l $. Also
$ \sigma_{l-k} > \sigma_{l-k+1} = \cdots = \sigma_{l} = 0$ if and only if  $dim\left( L \cap M\right) = k$.
\end{lemma}
Furthermore, if $\bP_L$ and $\bP_M$ represent the orthogonal projections onto the linear subspaces $L$ and $M$ respectively, then the cosines of the principal angles between $L$ and $M$ correspond to the square roots of the eigenvalues of $\bP_L\bP_M$. 

Consider now random samples of size $n$ from an elliptical $\mE(\bmu,\bGamma;G)$ distribution for which $\bmu = \bzero$ and for which $\bGamma$ has two distinct eigenvalues 
with multiplicities $d_1$ and $d_2 = d-d_1$. As noted previously, one can assume without loss of generality that the samples come from a multivariate normal distribution
$Normal_d(\bzero,\bGamma)$. Furthermore, by orthogonal and scale equivariance, one can also assume without loss of generality that $\bGamma$ is a diagonal matrix with the first $d_1$
diagonal elements equal to $\gamma > 1$ and the last $d_2$ elements equal to $1$, i.e.\
\begin{equation} \label{sim}
\bGamma =  diag(\underbrace{\gamma, \ldots, \gamma}_{d_{1}}, \underbrace{1, \ldots, 1}_{d-d_{1}})
\end{equation} 
The eigenprojection associated with the largest root of $\bGamma$, i.e.\  $\bP_1 $, then 
has its upper left $d_1 \times d_1$ block equal to the identity matrix of order $d_1$, with all the other blocks equal to zero. The principal angles between the space spanned by
an estimated eigenspace, say $\bhP_1$, and the space spanned by $\bP_1$ can then be computed by taking the arccosines of the square roots of the eigenvalue of the upper left
$d_1 \times d_1$ block of $\bhP_1$. For the case $d_1=1$, the sole principal angle is then the arcosine of the absolute value of the first element of the normalized eigenvector
estimate.

Simulations were undertaken in dimensions $d = 2, 3$ and $5$ for scatter structures of the form (\ref{sim}). Note that when $d=2$, the only possible case is $d_1 =1$, 
for which there is at most one non-zero principal angle. For $d=3$, the two possible cases are $d_1=1$ or $d_1 =2$, for which again there is only one non-zero principal angel. 
For $d=5$, the cases $d_1 = 1$  or $d_1 = 4$ have at most one non-zero principal angle, whereas the cases $d_1 = 2$ or $d_1 =3$ have at most two non-zero principal angles.  
For each of these cases, we consider three situations, namely when the major principal component space explains $90\%$, $95\%$ and 99\% of the total variance respectively.  
In two dimensions with $d_1 =1$, this corresponds to $\gamma = 9, 19$ and $99$.  In three dimensions, for $d_1 =1$ the corresponding values of $\gamma =18, 38$ and $198$, and 
for $d_1=2$ they are $\gamma = 4.5, 9.5$ and $49.5$. In five dimensions, for $d_1 =1$ the corresponding values of $\gamma = 36, 76,$ and $396$, for $d_1 = 2$ they are
$\gamma = 13.5, 28.5$ and $148.5$, for $d_1 =3$ they are $6, 12.67$ and  $66$ and for $d_1 = 4$ they are $\gamma = 2.25,  4.75$ and  $24.75$.

For the \emph{ASSCM}, due to its affine equivariance, it was only necessary to implement simulations for the case $\bGamma = \bI_{d}$.  For a fixed sample size and dimension, $10,000$ datasets were generated from a $Normal_d(\bzero,\bI_d)$ distribution.  The \emph{ASSCM}, $\bhT_n$, was calculated for each dataset.  By affine equivariance, the \emph{ASSCM} matrices for other scatter structures were obtained by transforming $\bhT_n \rightarrow \bGamma^{1/2}\bhT_n\bGamma^{1/2}$.  
For the \emph{SSCM}, since it is only scale and orthogonally equivariant, it was necessary to implement simulations for each sample size, dimension and scatter structure considered.  For each case considered, $10,000$ datasets were generated and the \emph{SSCM}, $\bhS_n$, was calculated for each dataset. The principal angles between $\bhP_{\mT,1}$ and $\bP_1$, as well as the
principal angles between $\bhP_{\mS,1}$ and $\bP_1$, were then calculated as previously described. Denote the principal angles between $\bhP_{\mT,1}$ and $\bP_1$ by
$\widehat{\tau}_1 \ge \ldots \ge \widehat{\tau}_{d_1} \ge 0$ and the principal angles between $\bhP_{\mS,1}$ and $\bP_1$ by 
$\widehat{\omega}_1 \ge \ldots \ge \widehat{\omega}_{d_1} \ge 0$. Since large deviations of the
principal angles from zero indicate a poor estimate, the sum of the squared principal angles gives a measure of loss of an eigenprojection estimate. 
A measure of relative efficiency of $\bhP_{\mS,1}$ to $\bhP_{\mT,1}$ can then be obtained by comparing the size of
$T_n = \widehat{\tau}_{1}^2 + \cdots + \widehat{\tau}_{d_{1}}^2$ to that of $\Omega_n = \widehat{\omega}_{1}^2 + \cdots + \widehat{\omega}_{d_{1}}^2$
using some measure of central tendency.  Here, we consider both the expected value and the median and so define the following two measures of finite sample relative
efficiency
\begin{equation} \label{arn}
RE_{1,n}\left[\bhP_{\mS,1}, \bhP_{\mT,1}; \bGamma \right] = \frac{E\left[T_n\right]}{E\left[\Omega_n\right]} \quad \mbox{and} \quad
RE_{2,n}\left[\bhP_{\mS,1}, \bhP_{\mT,1}; \bGamma \right] = \frac{median\left[T_n\right]}{median\left[\Omega_n\right]}
\end{equation}
We show in Appendix \ref{app-arn} that the limiting value of $RE_{2,n}$ is equal to the asymptotic relative efficiency given by (\ref{are}), where here
$\rho^2 = \lambda_{(2)}/\lambda_{(1)} = 1/\gamma$, and we conjecture that this also holds for $RE_{1,n}$.

\begin{figure}[t] \label{sim2} \hspace*{-.3cm}
\begin{tabular}{ccc}
\includegraphics[scale= 0.35]{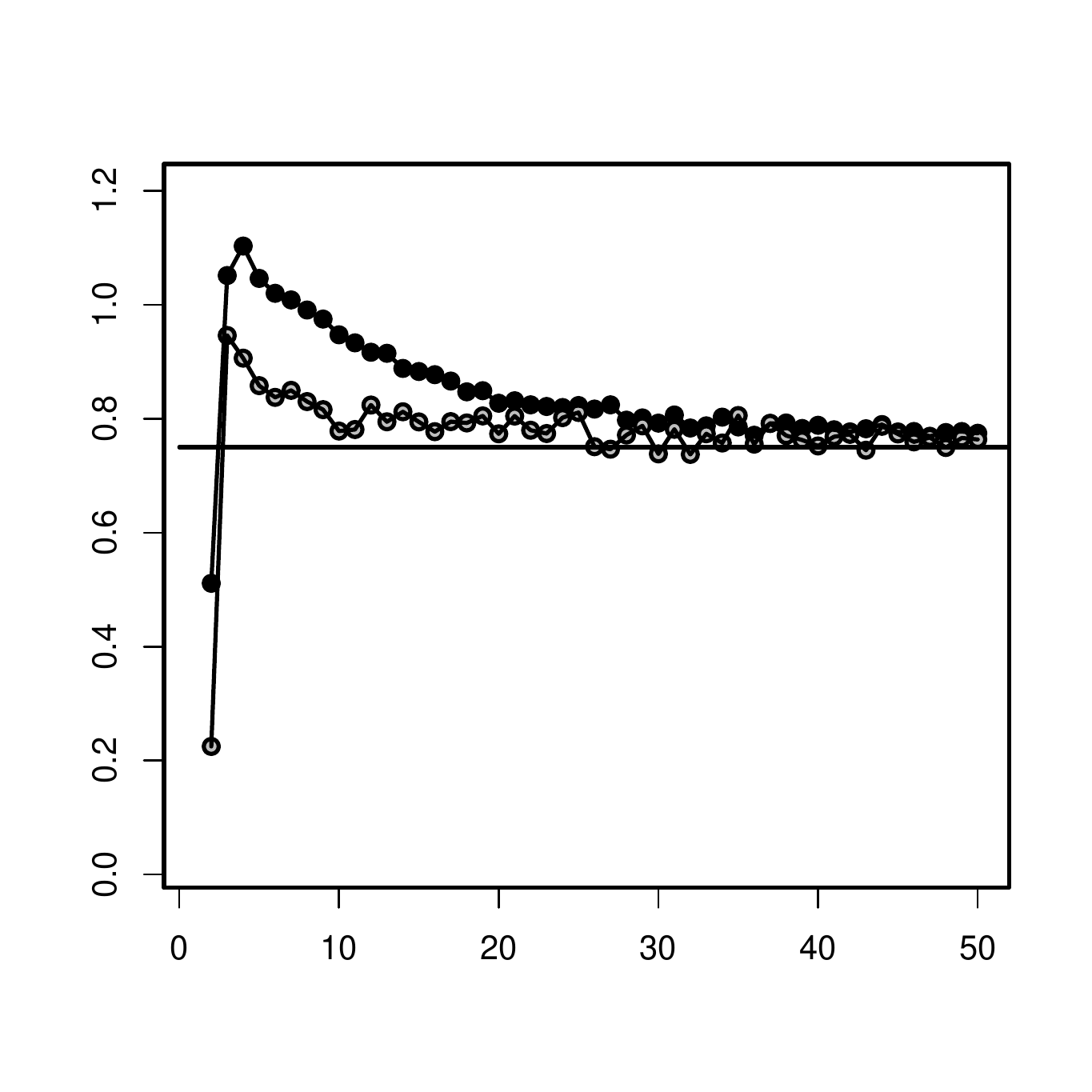}  &   \includegraphics[scale= 0.35]{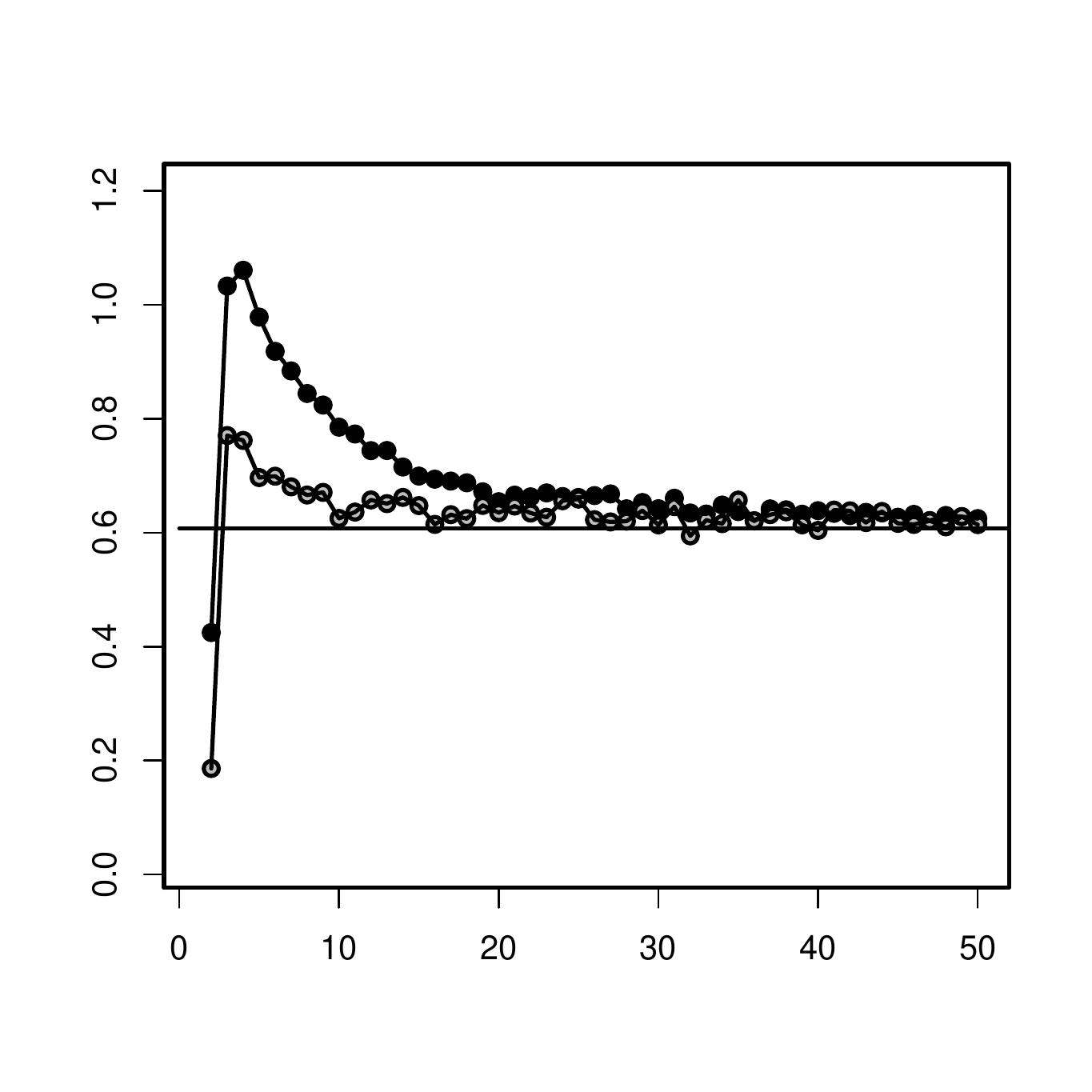}   &  \includegraphics[scale= 0.35]{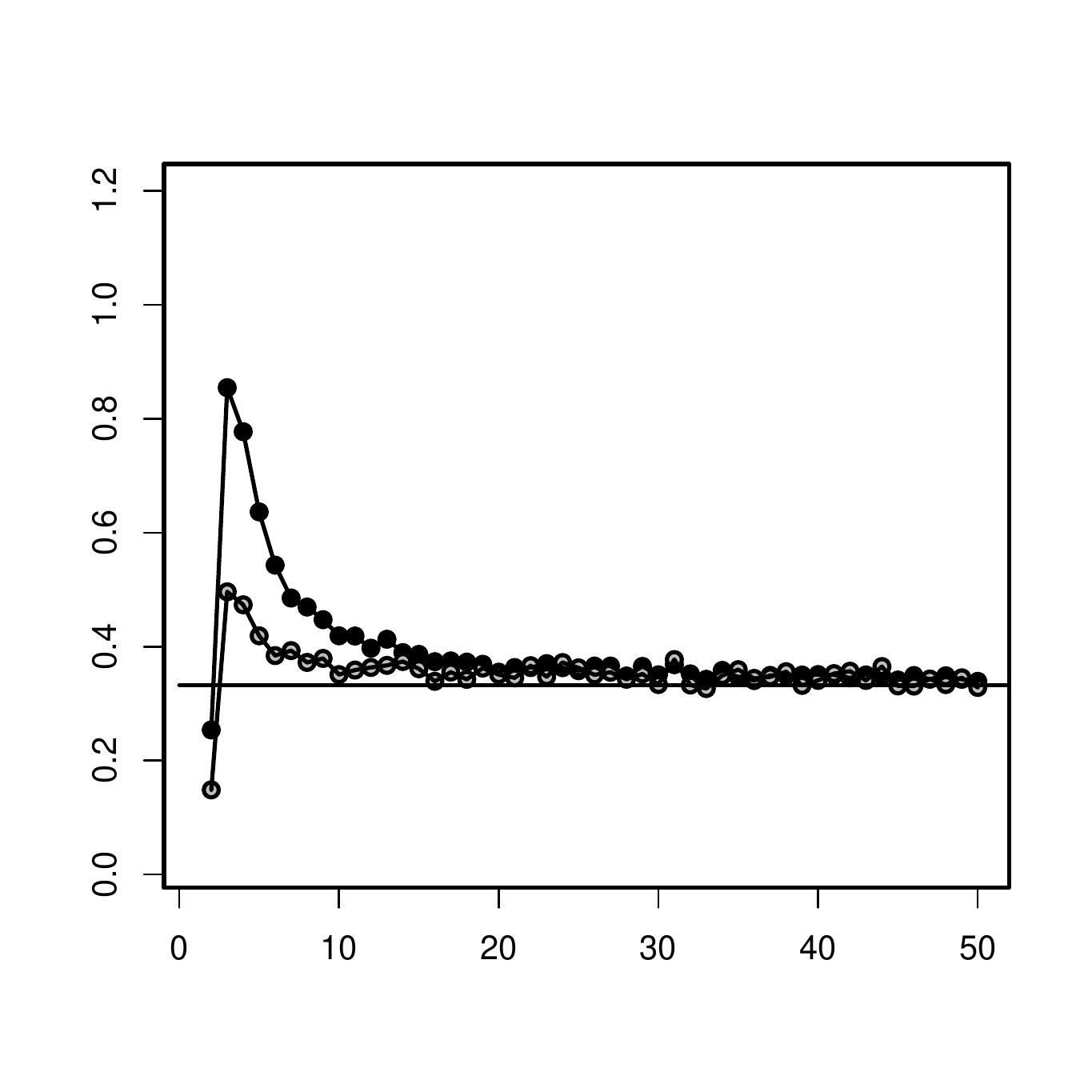} \\[-16pt]
$d_1 =1, \gamma = 9.0$  & $d_1 =1, \gamma = 19.0$ & $d_1 =1, \gamma = 99.0$
\end{tabular}
    \caption{Finite sample relative efficiencies in $\mathbb{R}^{2}$ for the scatter structure (\ref{sim}), as a function of sample size, of the 
		eigenprojection of the \emph{SSCM} to that of  the \emph{ASSCM}. Black and grey circles refer to $RE_{1,n}$ and $RE_{2,n}$, (\ref{arn}) respectively. }
\end{figure}

Figures 4, 5, and 6 show the simulated values of the relative efficiencies (\ref{arn}) for dimensions $d = 2, 3$ and $5$ respectively. For $d = 2$ and $3$, the sample 
sizes are $n = d, \ldots, 50$. When $d=5$, the simulations were done for sample sizes  $n = 5, 10, \ldots, 125$.  
In all of the figures, it can be observed that the finite sample relative efficiency quickly decreases to the asymptotic relative efficiency, which
is indicated by the horizontal line. For smaller sample sizes, the finite sample relative efficiency using the median based measure, $RE_{2,n}$, is considerably smaller
than that using the expected value based measure, $RE_{1,n}$. Also, for relatively small samples sizes, the deficiency of the 
\emph{SSCM} is not as pronounced as in larger sample sizes. The one exception to this trend is when $n =d$. For this case, it is known that the
\emph{ASCCM} or Tyler's matrix is proportional to the sample covariance matrix, and that the sample covariance matrix is proportional to the maximum likelihood estimate for
the shape matrix $\bGamma$ not only under the multivariate normal model but also under any elliptical model, see \citep{Tyler10}.

\begin{figure}[h] \label{sim3} \hspace*{-.3cm}
\begin{tabular}{ccc}
	 \includegraphics[scale= 0.35]{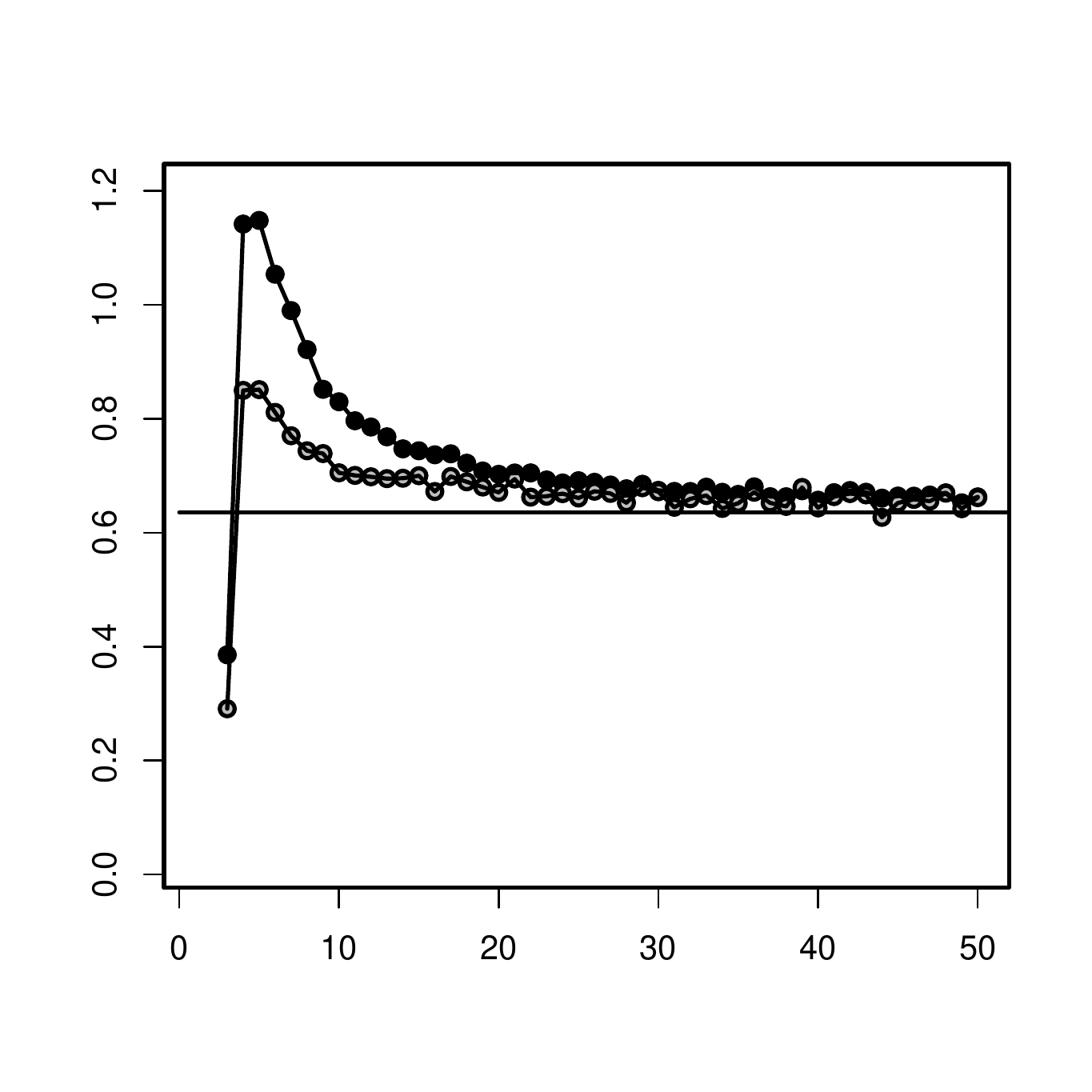}  &   \includegraphics[scale= 0.35]{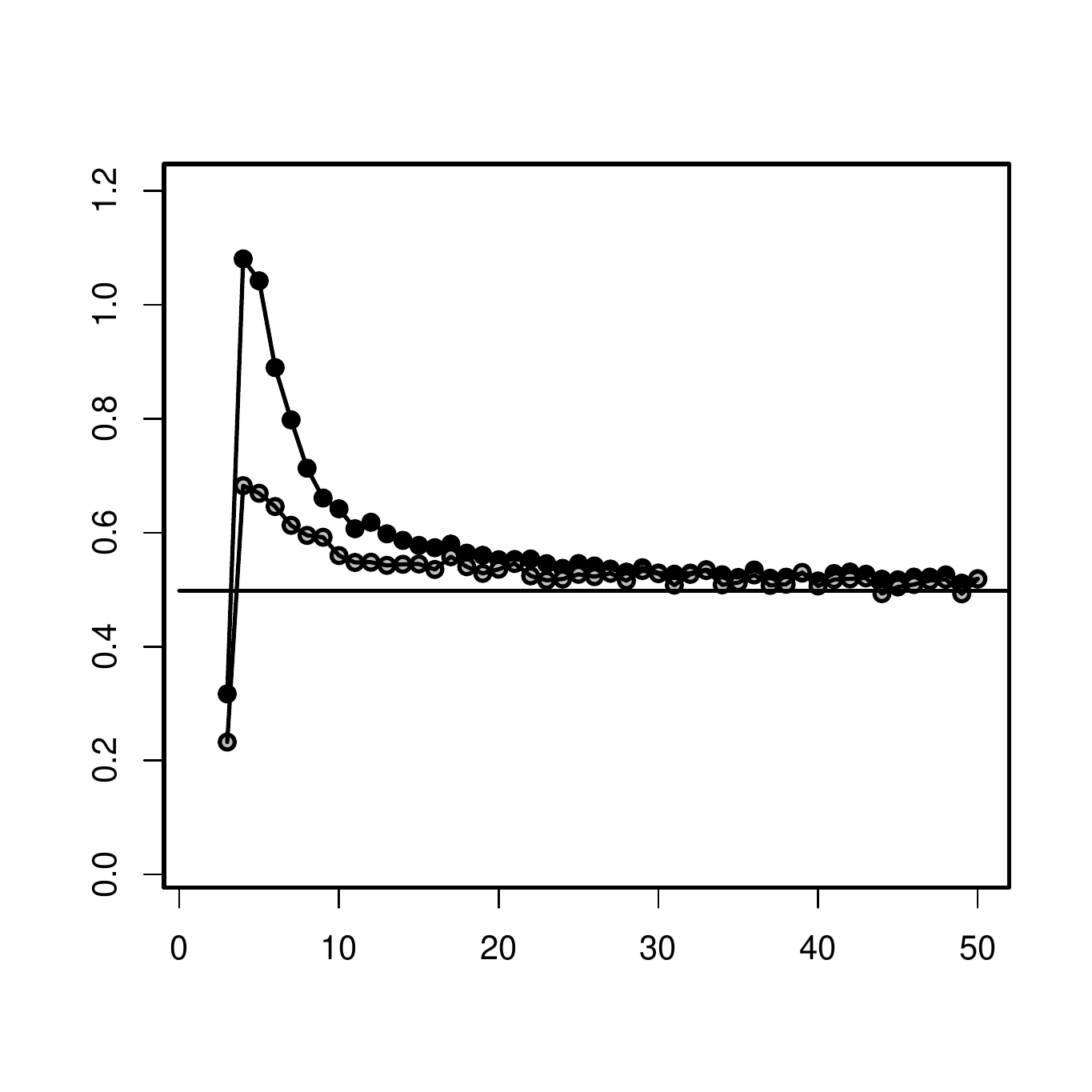}   &  \includegraphics[scale= 0.35]{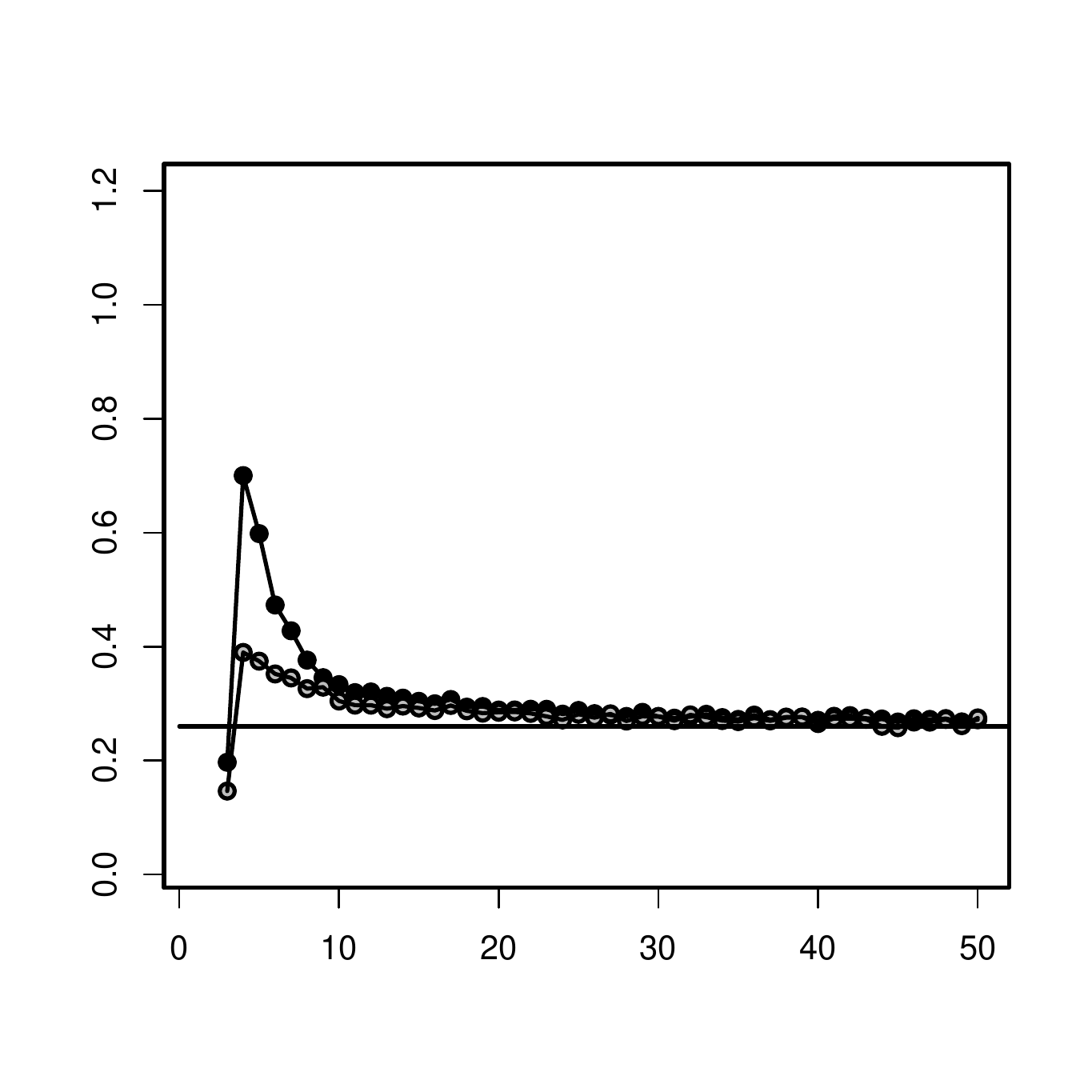} \\[-16pt]
	$d_1 =1, \gamma = 18.0$  & $d_1 =1, \gamma =38.0$ & $d_1 =1, \gamma = 198.0$ \\
  \includegraphics[scale= 0.35]{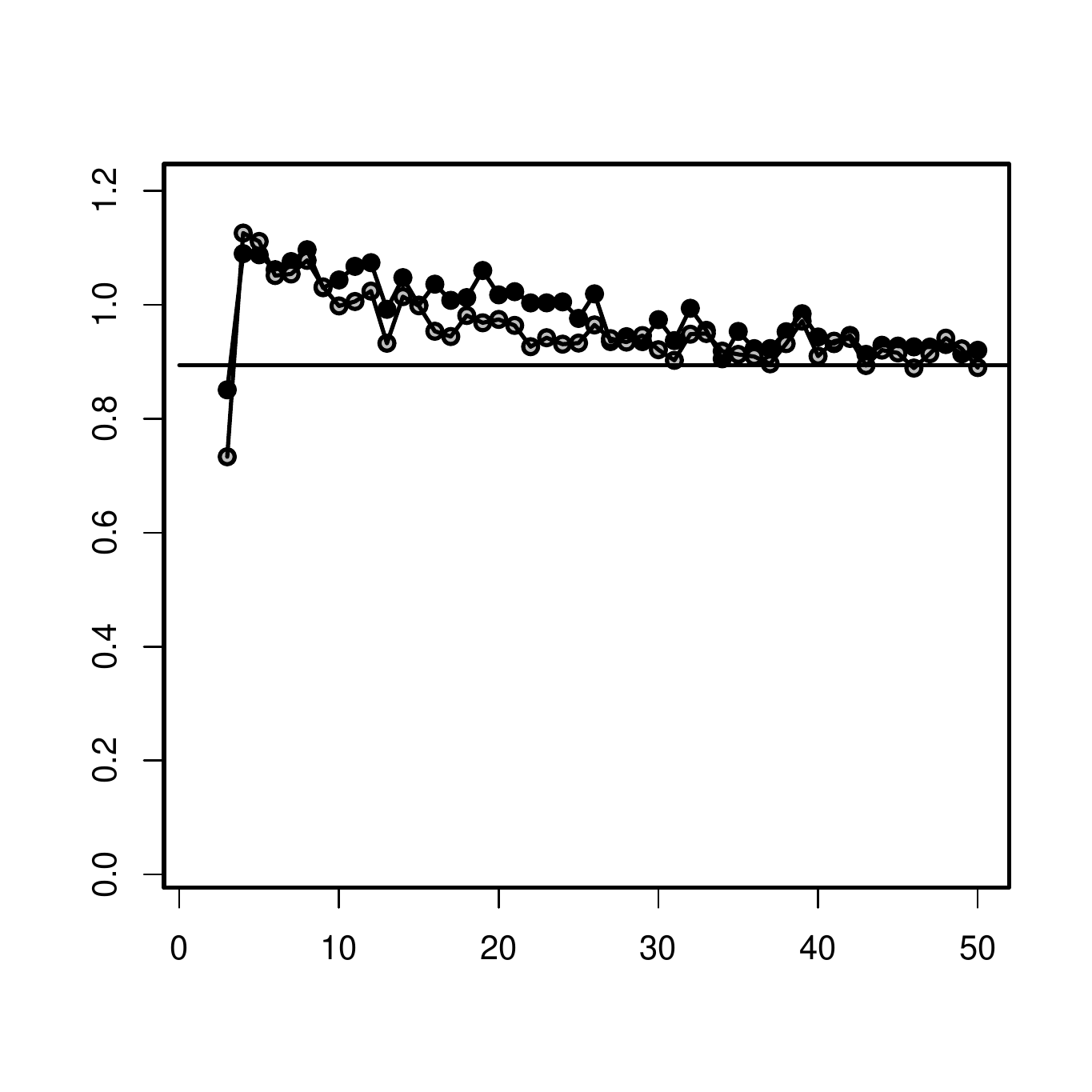}  &   \includegraphics[scale= 0.35]{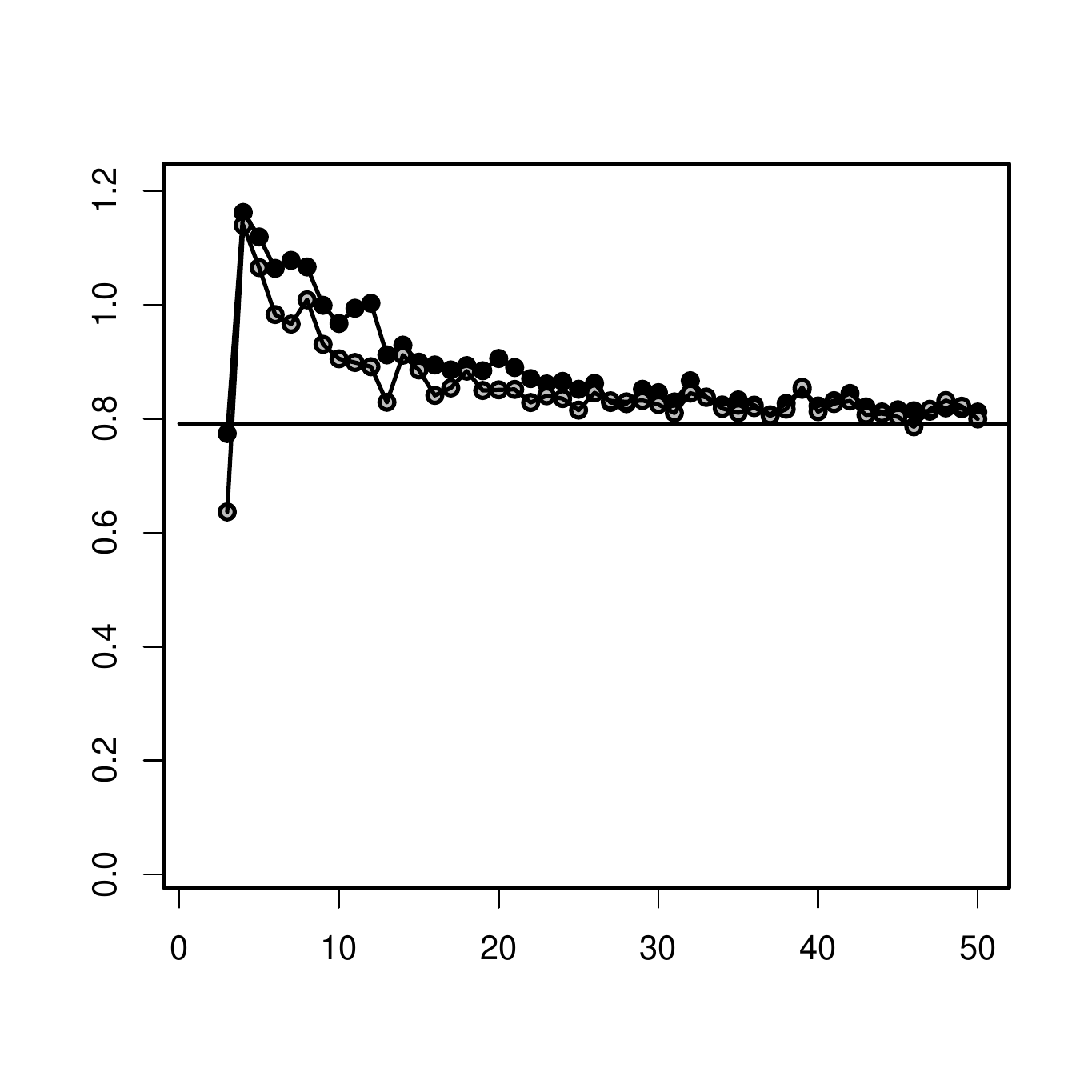}  &   \includegraphics[scale= 0.35]{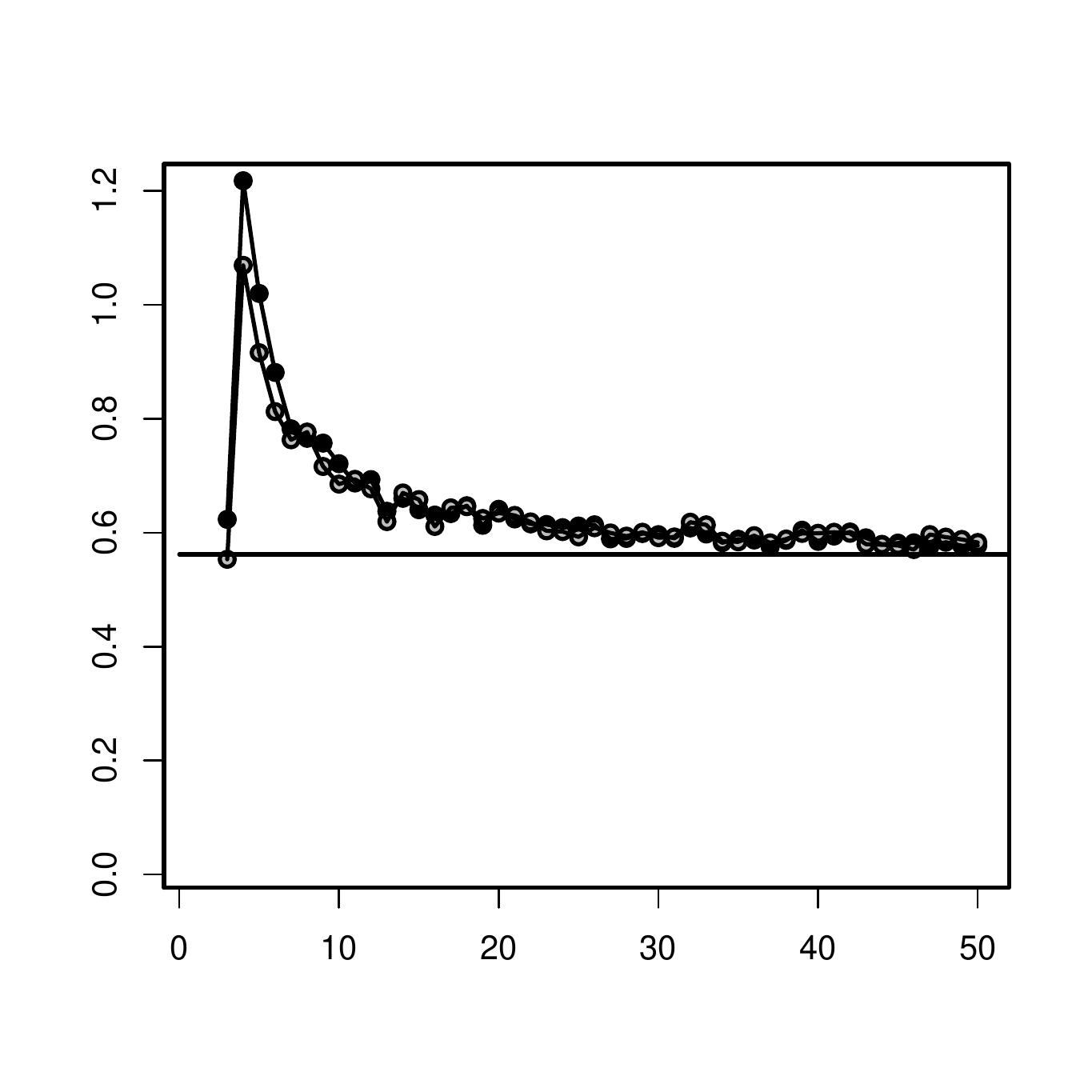} \\[-16pt]
	$d_1 =2, \gamma = 4.5$  & $d_1 =2, \gamma =9.5$ & $d_1 =2, \gamma = 49.5$
\end{tabular}
    \caption{Finite sample relative efficiencies in $\mathbb{R}^{3}$ for the scatter structure (\ref{sim}), as a function of sample size, of the 
		eigenprojection of the \emph{SSCM} to that of  the \emph{ASSCM}. Black and grey circles refer to $RE_{1,n}$ and $RE_{2,n}$, (\ref{arn}) respectively. }
\end{figure}

\begin{figure}[!htbp] \label{sim5} \hspace*{-.3cm}
\begin{tabular}{ccc}
	  \includegraphics[scale= 0.35]{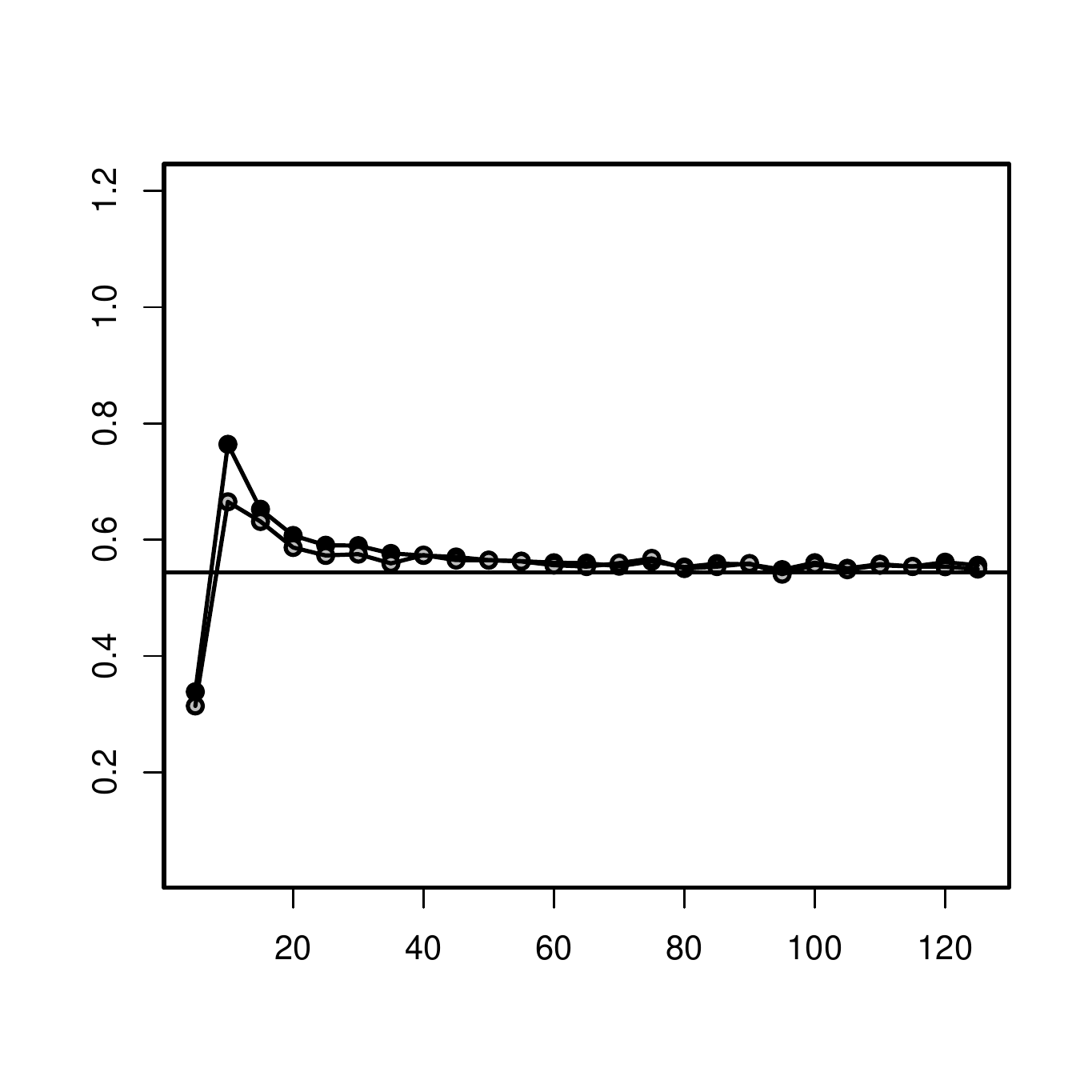}   &  \includegraphics[scale= 0.35]{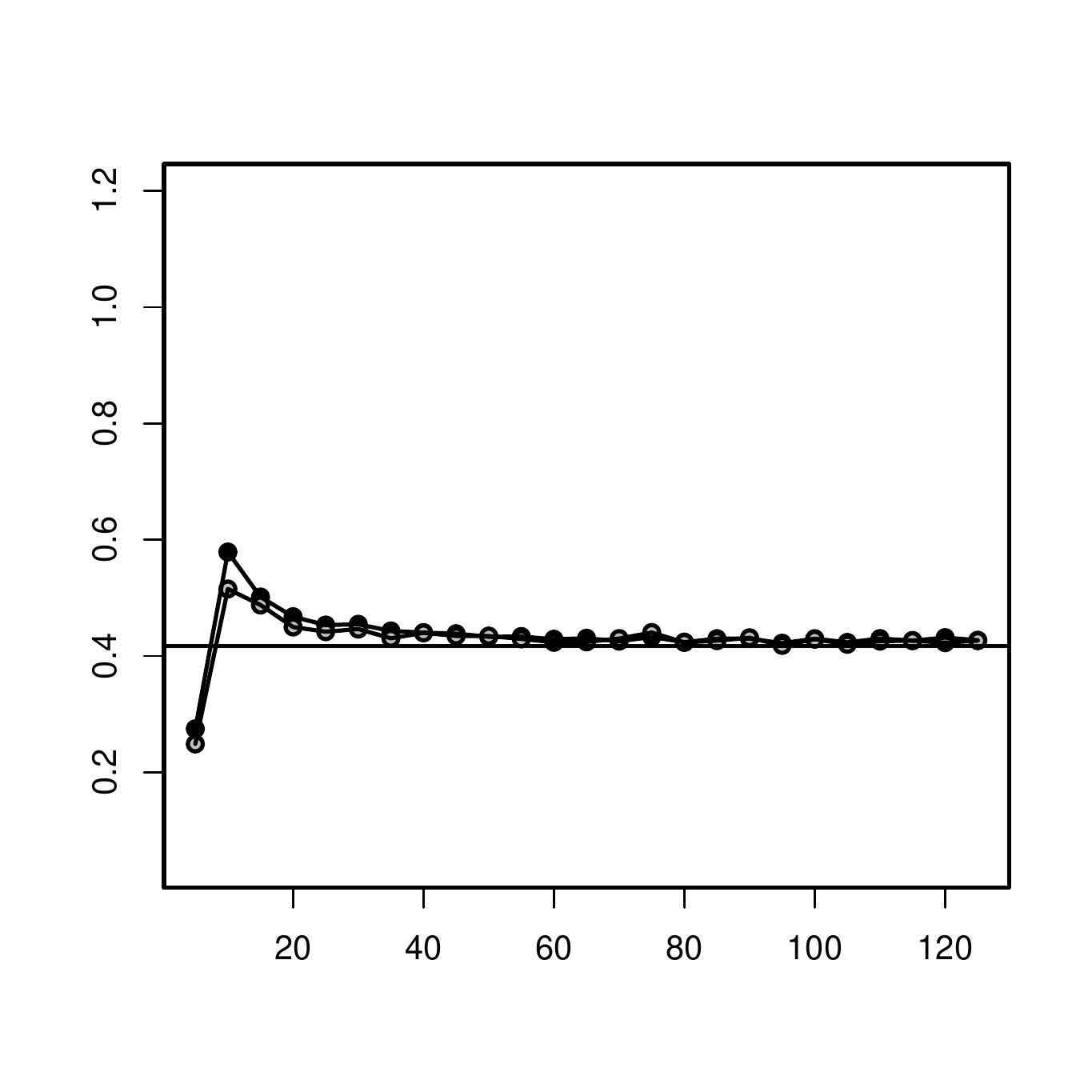}   &  \includegraphics[scale= 0.35]{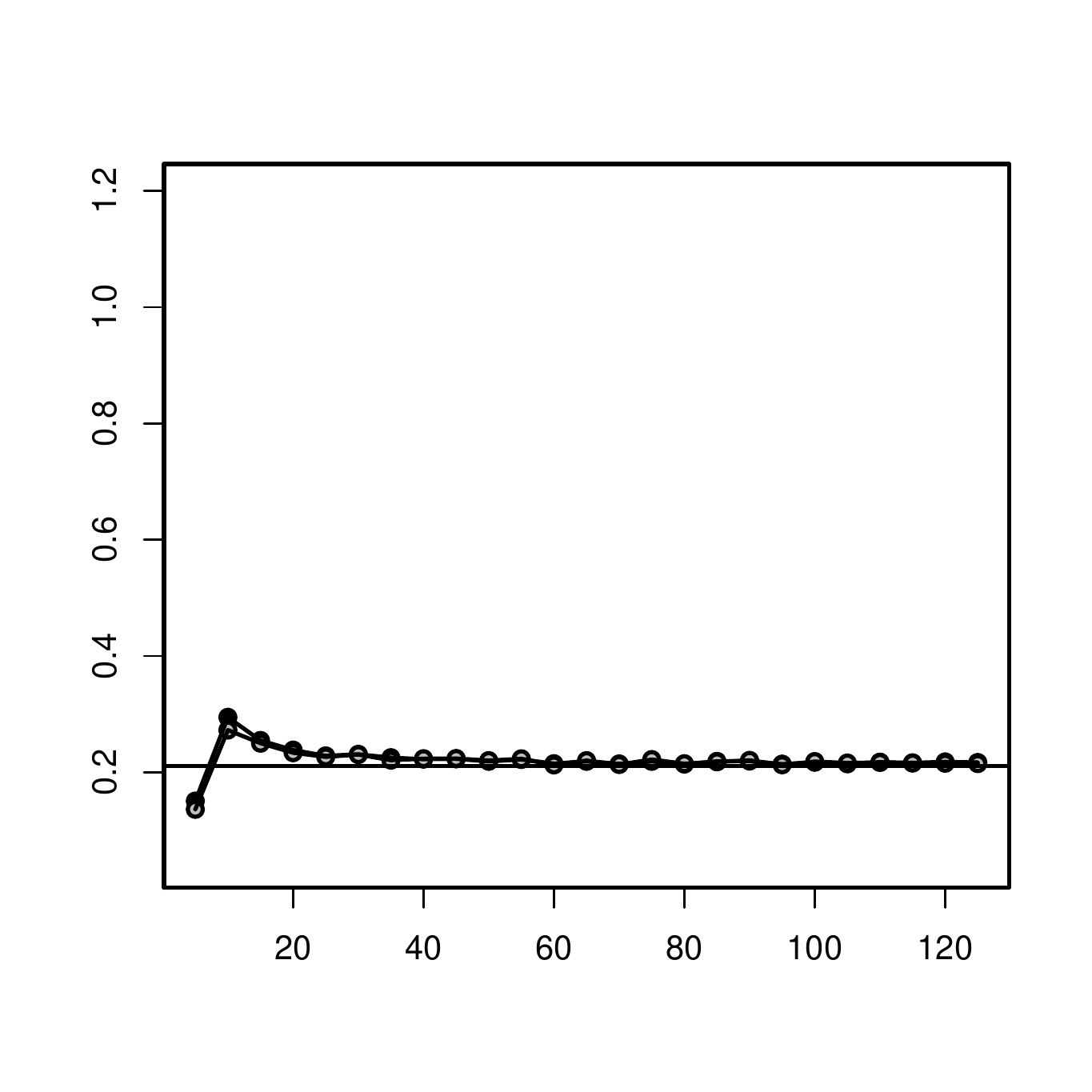} \\[-16pt]
		$d_1 =1, \gamma = 36.0$  & $d_1 =1, \gamma = 76.0$ & $d_1 =1, \gamma = 396.0$ \\	
  \includegraphics[scale= 0.35]{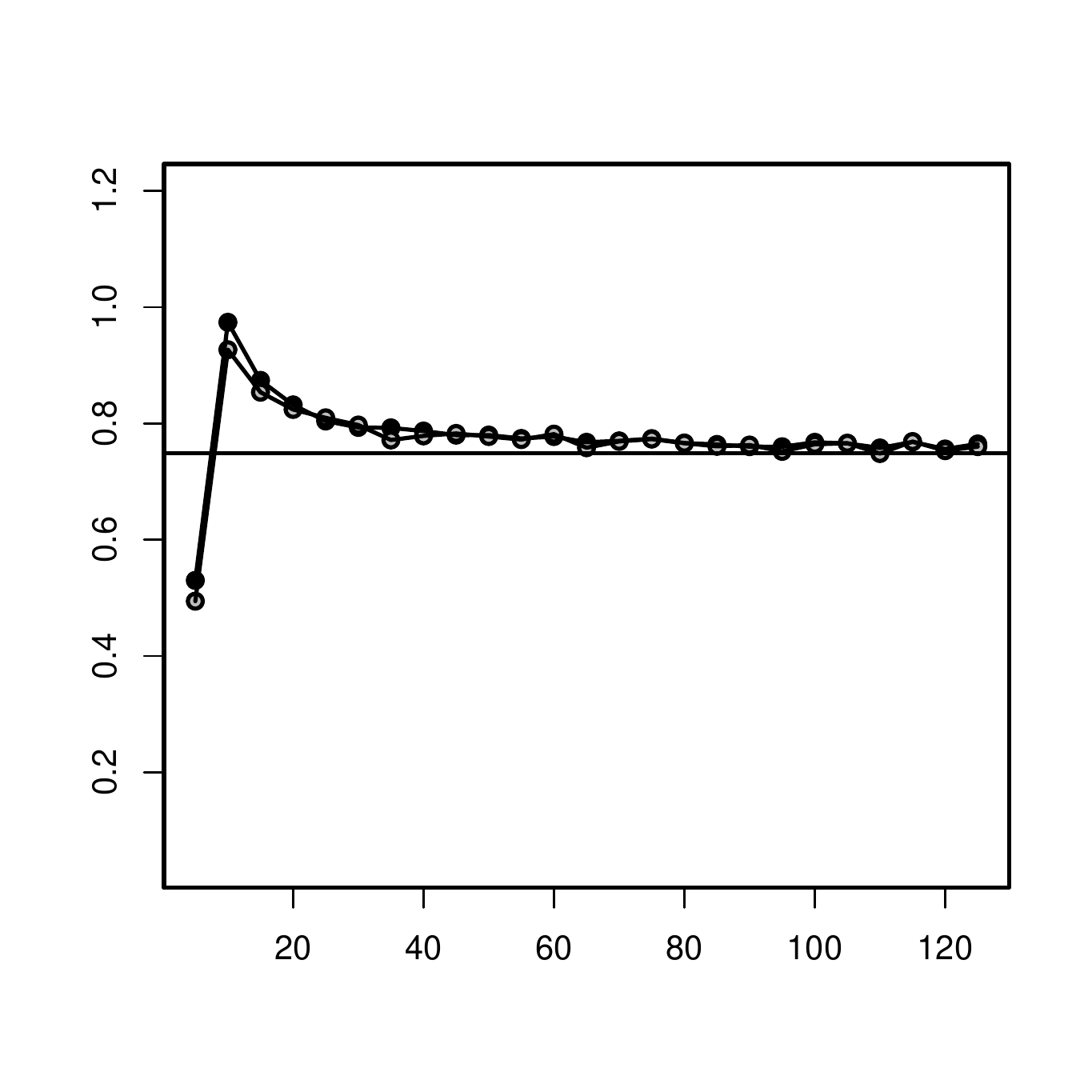}   &  \includegraphics[scale= 0.35]{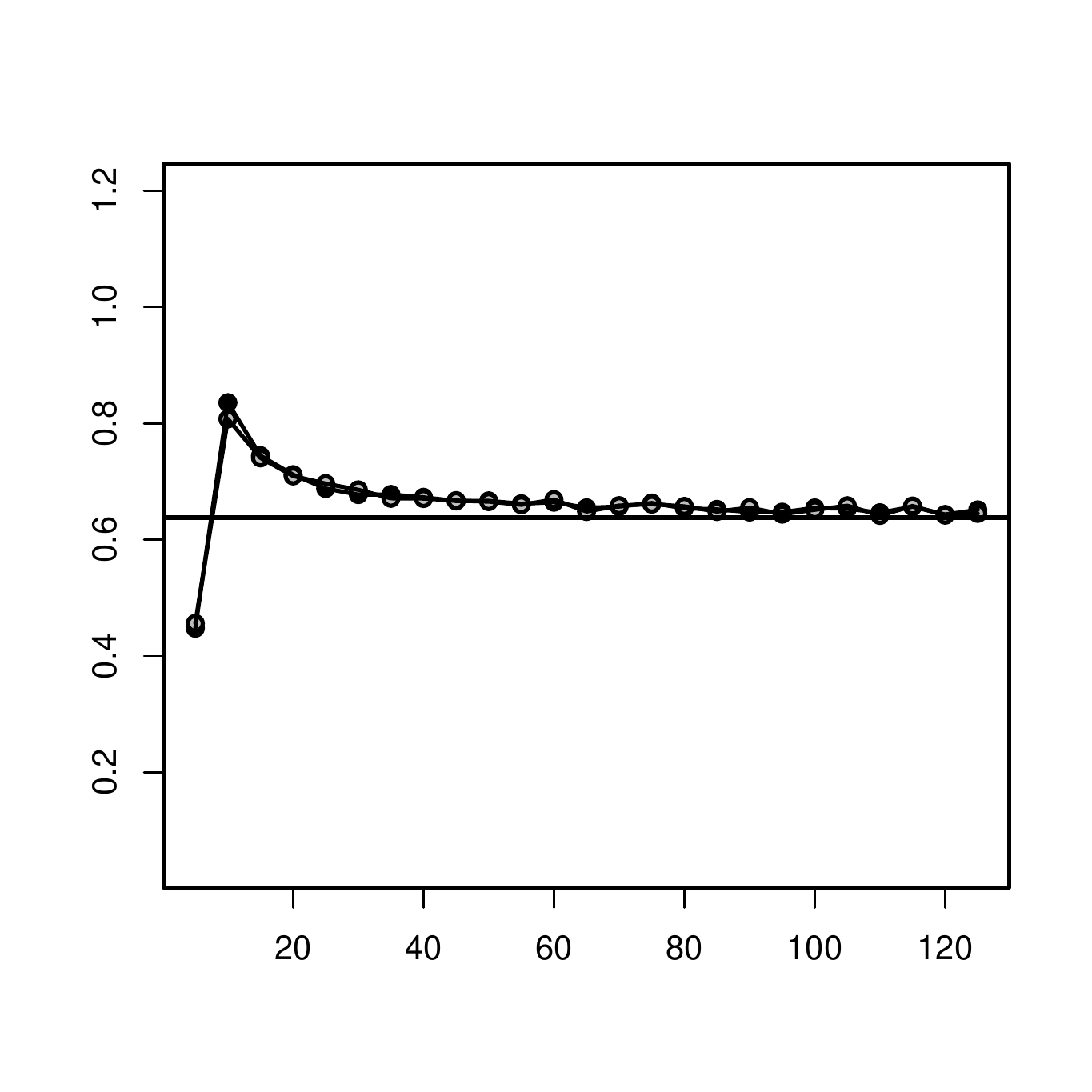}  &   \includegraphics[scale= 0.35]{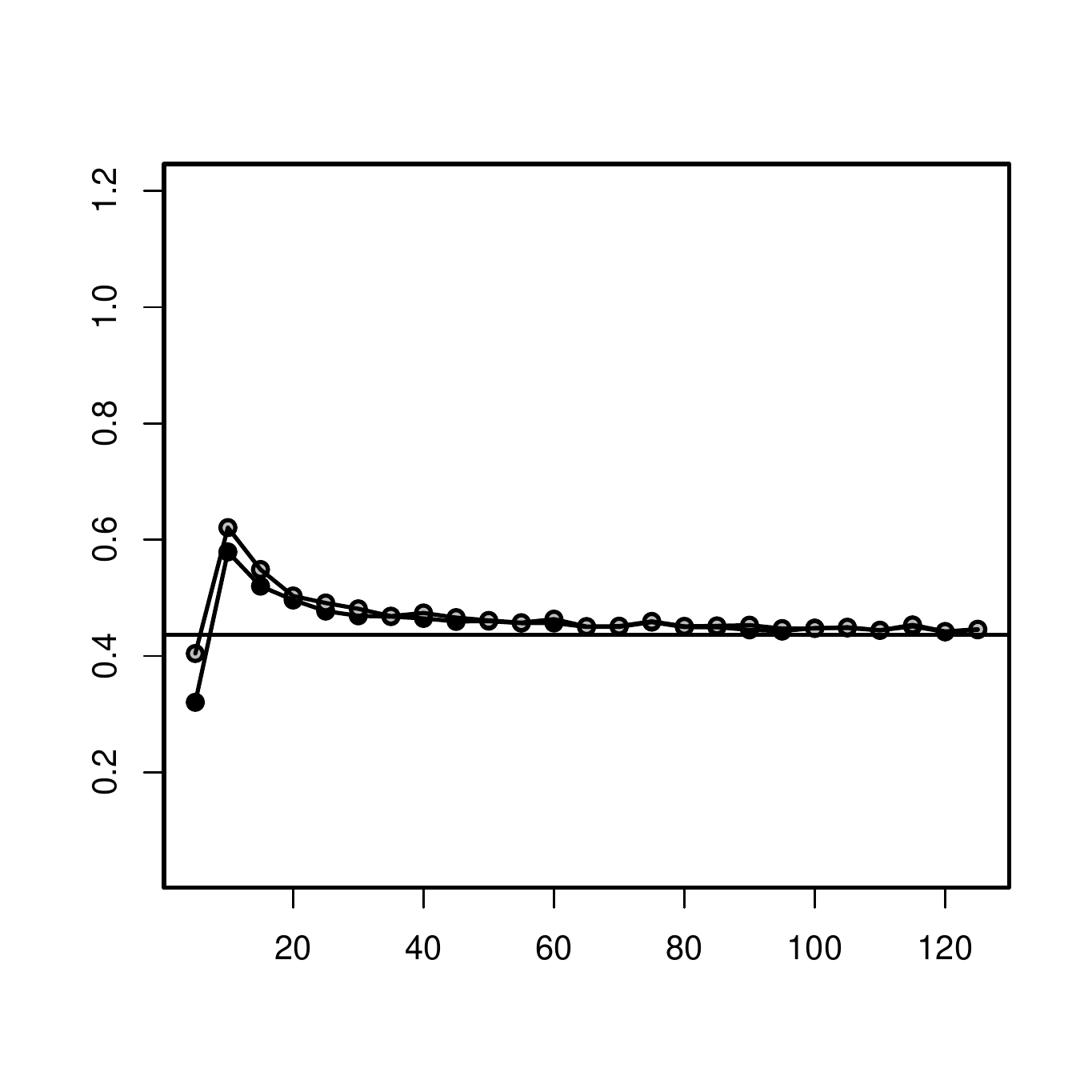} \\[-16pt]
		$d_1 =2, \gamma = 13.5$  & $d_1 =2, \gamma =28.5$ & $d_1 =2, \gamma = 148.5$ \\
	  \includegraphics[scale= 0.35]{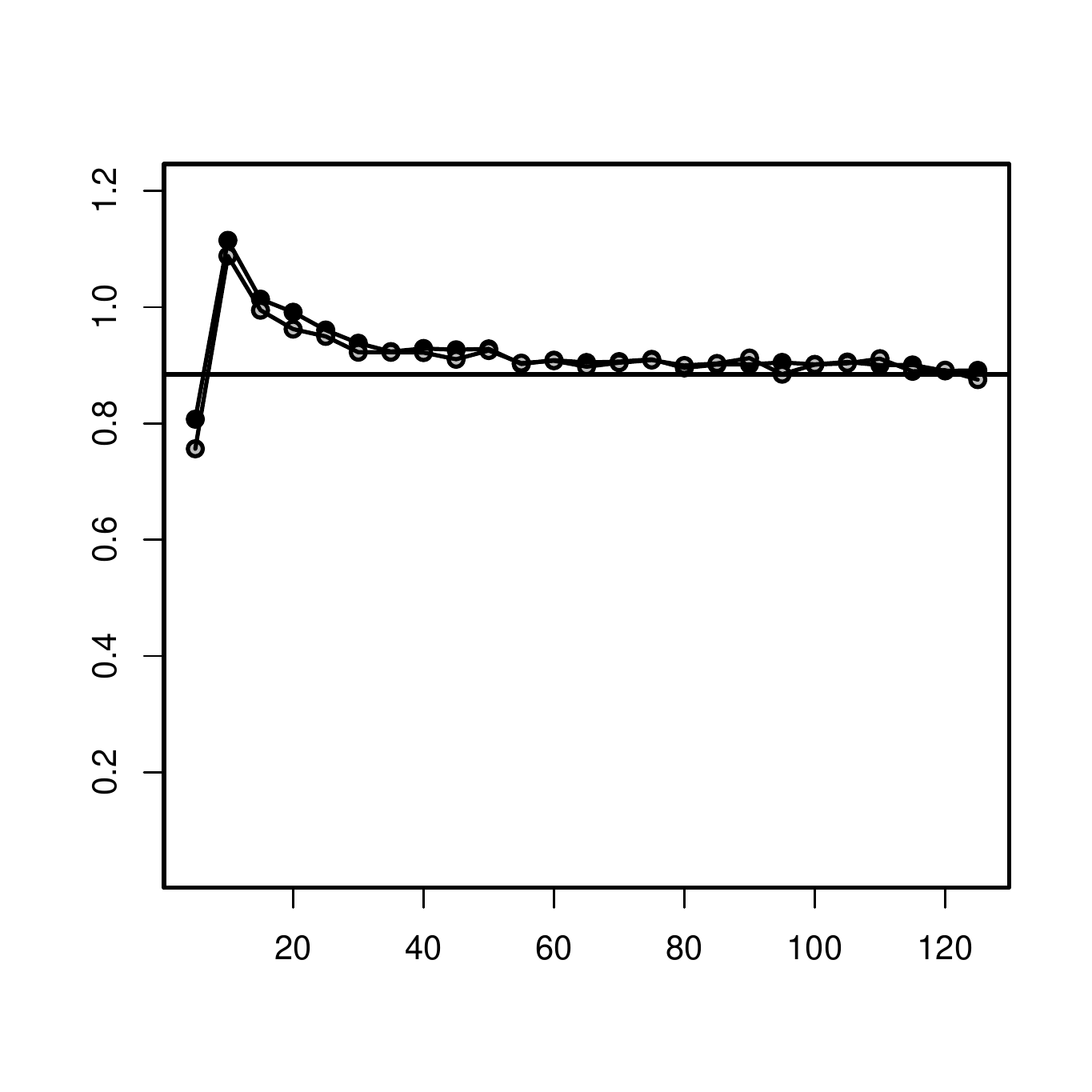}   &  \includegraphics[scale= 0.35]{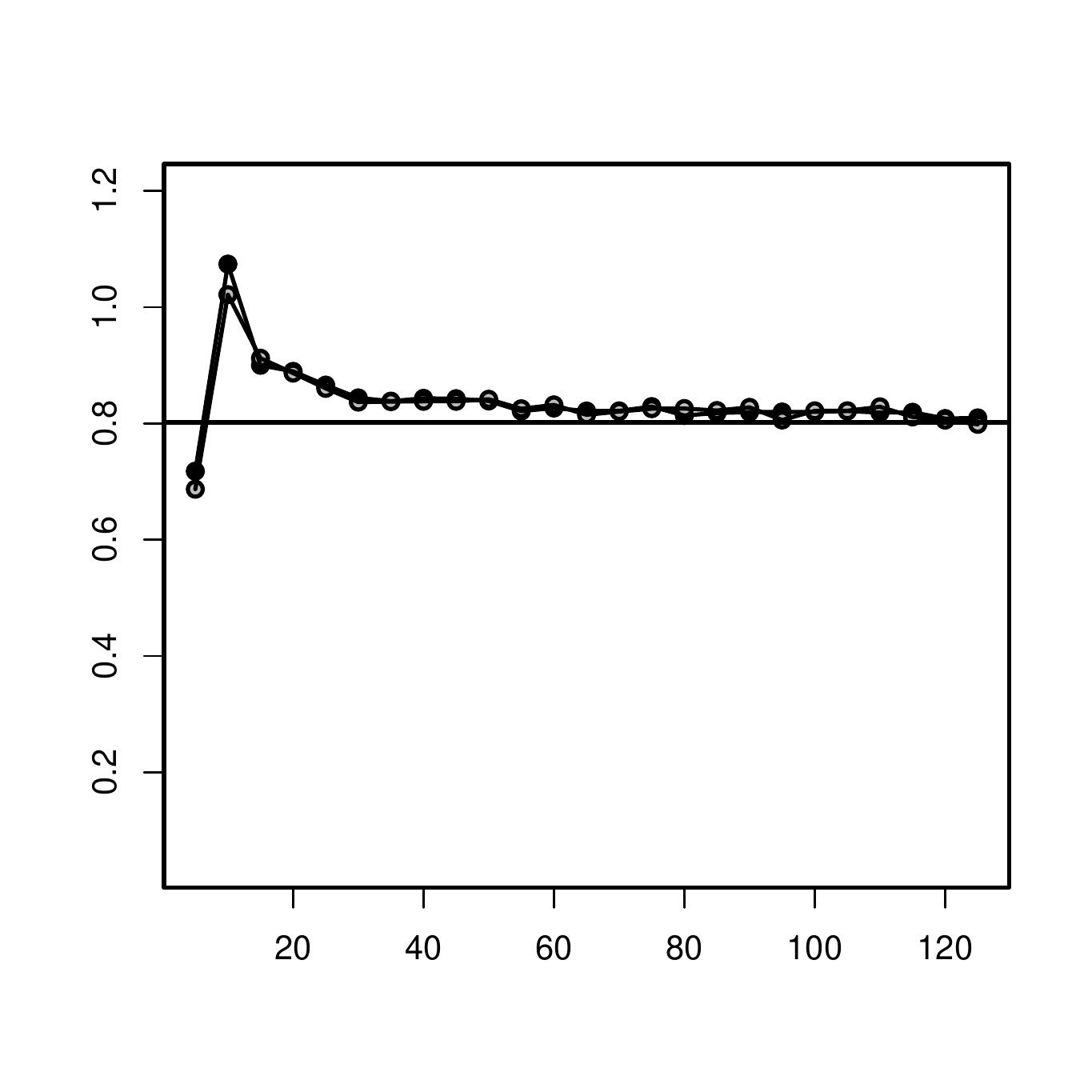}   &  \includegraphics[scale= 0.35]{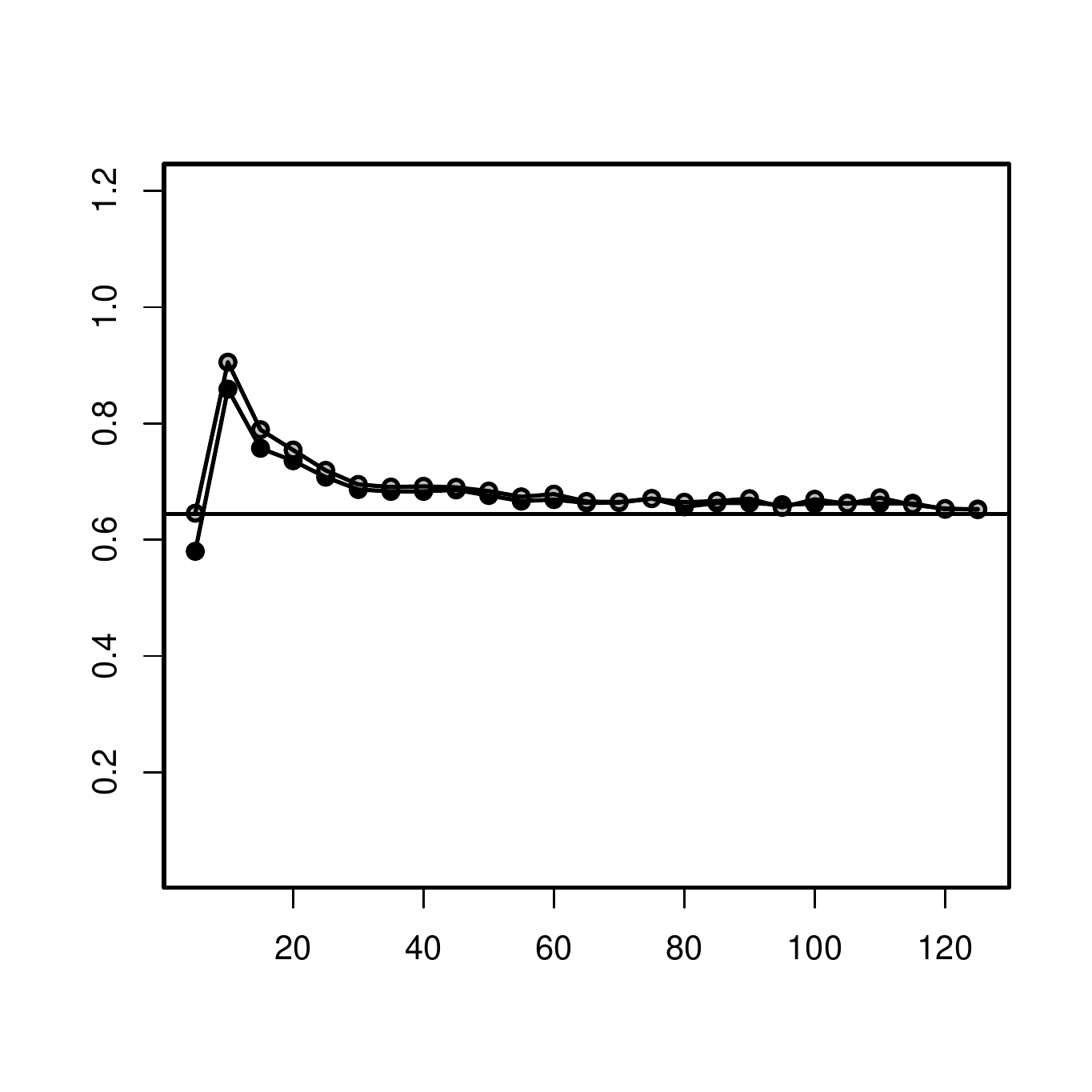} \\[-16pt] 
		$d_1 =3, \gamma = 6.0$  & $d_1 =3, \gamma =12.67$ & $d_1 =3, \gamma = 66.0$	\\
	 \includegraphics[scale= 0.35]{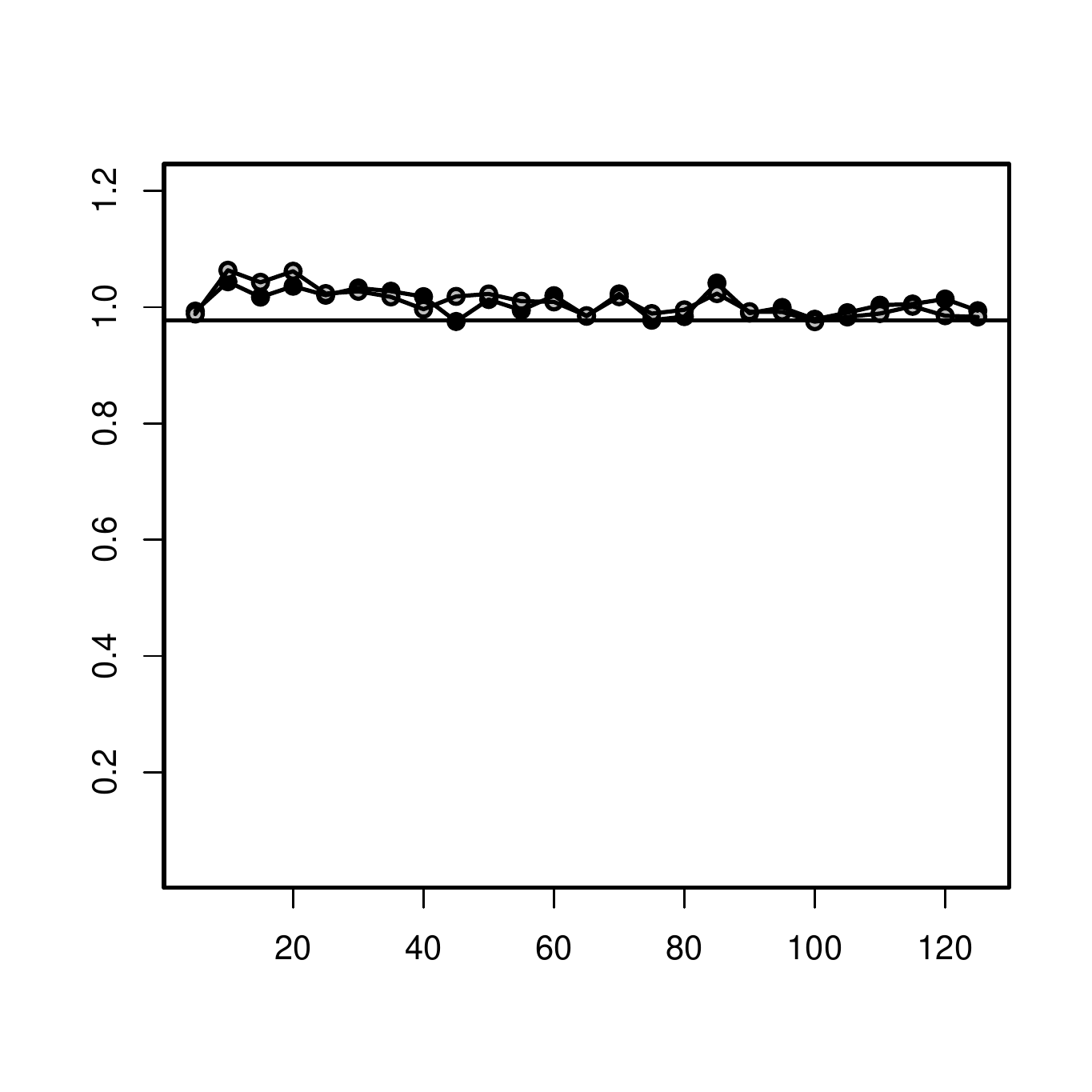}  &   \includegraphics[scale= 0.35]{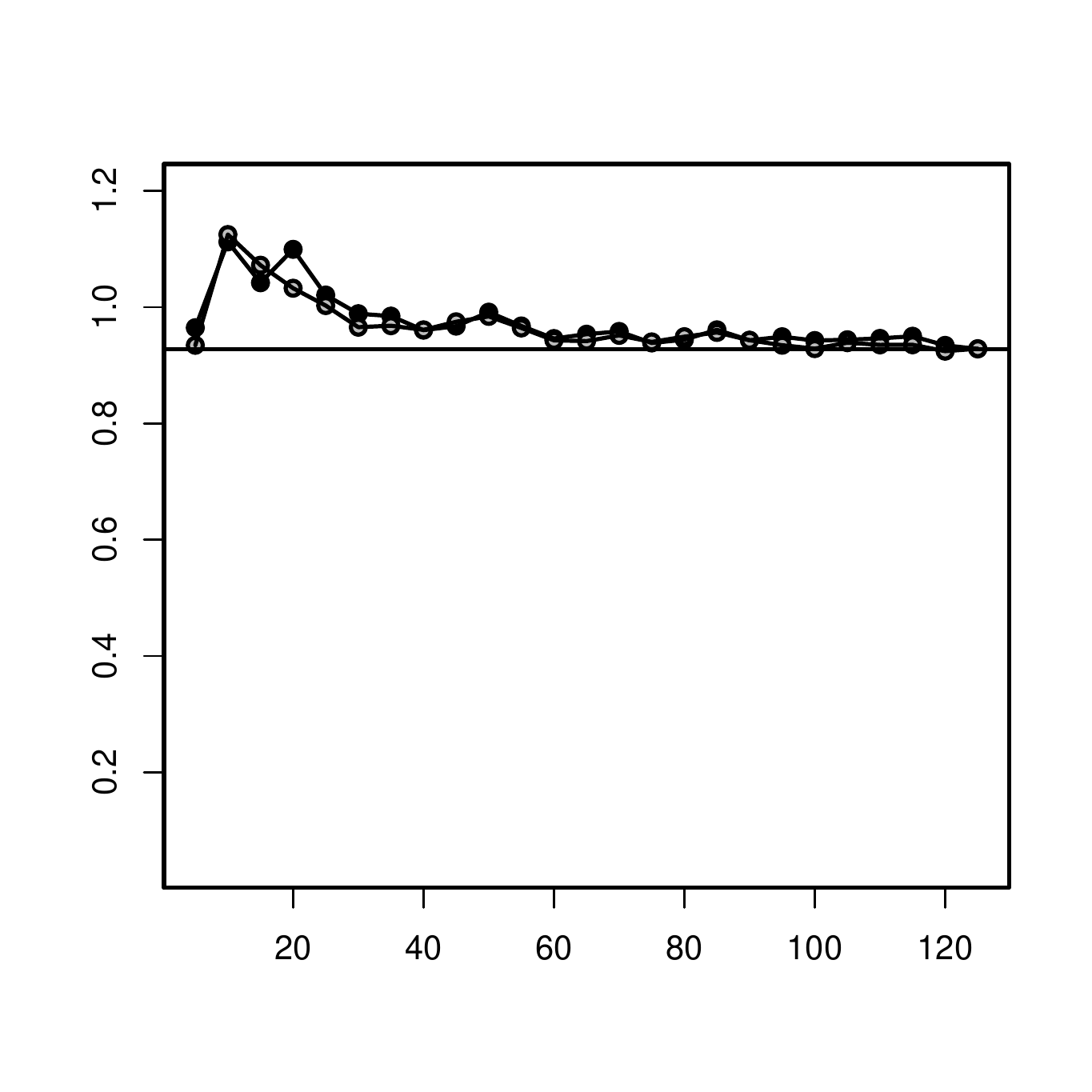}   &  \includegraphics[scale= 0.35]{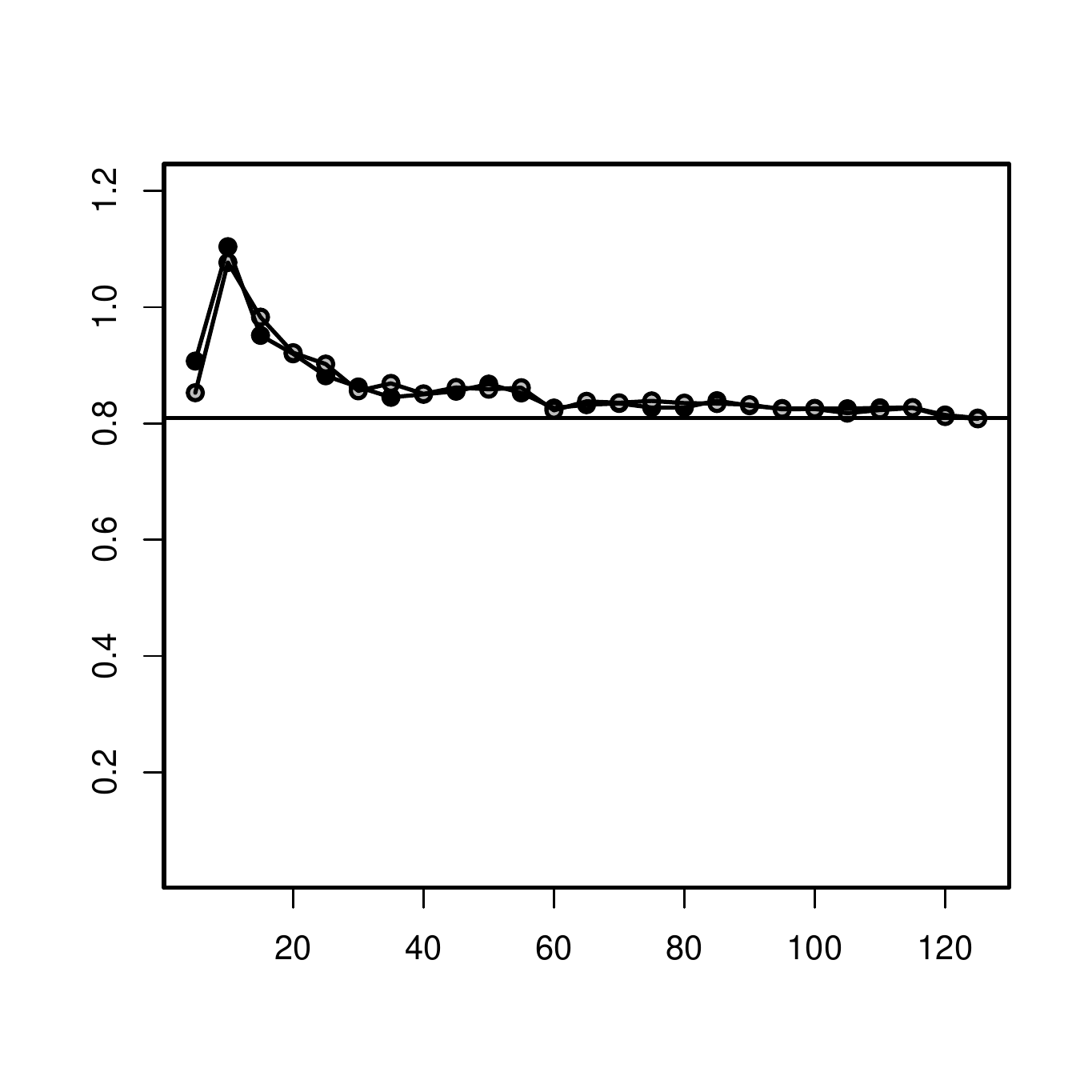} 	\\[-16pt]	
	 	$d_1 =4, \gamma = 2.25$  & $d_1 =4, \gamma = 4.75$ & $d_1 =4, \gamma = 24.75$
	\end{tabular}
    \caption{Finite sample relative efficiencies in $\mathbb{R}^{5}$ for the scatter structure (\ref{sim}), as a function of sample size, of the 
		eigenprojection of the \emph{SSCM} to that of  the \emph{ASSCM}. Black and grey circles refer to $RE_{1,n}$ and $RE_{2,n}$, (\ref{arn}) respectively. }
\end{figure}

\section{Concluding Remarks} \label{conclude}
The spatial sign covariance matrix (\emph{SSCM}) has been proposed in the past as an alternative estimate of scatter/shape based on its robustness 
properties and computational simplicity. These benefits, though, do not arise without a cost.  
For the case when the data arises from an elliptically symmetric distribution, it is known that in 
general the \emph{SSCM} is a biased estimate of shape parameter $\bGamma$, and in particular the eigenvalues of 
the SSCM are biased estimates of the eigenvalues of $\bGamma$.  However, as a 
consequence of the \emph{SSCM} being orthogonally equivariant, the corresponding eigenvectors, or more generally eigenprojections, 
of the SSCM are unbiased estimates of the eigenvectors/eigenprojections of $\bGamma$.  Due to its robustness, computational simplicity
and orthogonal equivariance,  the \emph{SSCM} has been suggested for extracting principal components vectors, which is
an orthogonally equivariant procedure.  However, this paper demonstrates that under the elliptical model, even the estimates of 
eigenvectors/eigenprojections based on the \emph{SSCM} are affected by the lack of affine equivariance of the \emph{SSCM} and
are highly inefficient in situations for which \emph{PCA} is of most interest.  

In the current paper it is proven that a consistency adjusted SSCM (obtain by replacing the 
eigenvalues of the SSCM estimate with unbiased ones) is asymptotically inadmissible when 
compared to a well studied affine equivariant version of the SSCM, Tyler's estimate of scatter, which
is referred to in the paper as the \emph{ASSCM}. The asymptotic relative efficiency of an
eigenprojection based on the \emph{SSCM} relative to that based on the \emph{ASSCM} is shown to be
quit severe, in particular for nearly singular scatter structures when the dimension of the principal component space 
with the highest variability is small compared to the dimension of the data. This situation is when \emph{PCA} is of  
most value as a dimension reduction method.  Finite sample simulations were implemented 
which also demonstrate the inefficiency of the eigenprojection estimates based on the \emph{SSCM} for relatively
small sample sizes. It should be noted that the current paper can be considered as complementary to \cite{Magyar12} in which the 
inefficiency of the spatial median, a popular orthogonally equivariant estimate of location was studied in detail under the 
elliptical model.  

As noted at the end of section \ref{Hyper}, the asymptotic efficiency of the \emph{SSCM} relative to any affine equaivariant
estimate can be obtain by making a simple scalar adjustment, not dependent on $\bGamma$, to the asymptotic relative efficiency
of the \emph{SSCM} to the \emph{ASSCM}. The main advantage of the \emph{SSCM} over a high breakdown point affine equivariant
scatter estimate is its computational simplicity.  The main advantage of the \emph{SSCM} over computationally simple affine
equivariant scatter estimates, such as the \emph{M}-estimates of scatter, is its higher breakdown point.  The \emph{SSCM}
has a breakdown point of $1/2$, whereas the \emph{M}-estimates of scatter have breakdown points at best $1/d$, with Tyler's
estimate of scatter acheiving a breakdown point of $1/2$, see e.g. \cite{Dumbgen05} who also show that the breakdown of the
\emph{M}-estimates occur only under very specific types of contamination. It is also worth noting, as argued in
\cite{Davies05}, that the concept of the breakdown point of a scatter matrix is not as meaningful outside of the 
affine equivariant setting. For example, the sample covariance matrix $\bS_n$ can be easily adjusted to have breakdown 
point $1/2$ by expressing it in terms of it spectral value decomposition $\bS_n = \bQ_n\bDelta_n\bQ_n^T$ and then
defining $\bS_n^* = \bQ_n\bDelta^*_n\bQ_n^T$, where $\bDelta^*_n$ is a diagonal matrix consisting of the squares of the marginal
median absolute deviations of $\bQ_n^T\bx_i$. Note that $\bS_n$ and $\bS^*_n$ have the same eigenvectors, but that $\bS^*_n$  is
only orthogonally equivariant and has breakdown point $1/2$. Finally, it should be noted that for affine equivariant scatter
estimates the concept of the breakdown point and the breakdown point at the edge are equivalent, but this is not the case for
non-affine equivariant scatter statistics, see \citet{Stahel81, Hampel86}. The breakdown point at the edge of a scatter statistic
is the multivariate equivalent of the \emph{exact fit} property in regression. It roughly corresponds to the proportion of contamination 
one needs to add to a lower dimensional data set in order for the scatter statistics to become non-singular. If the scatter statistic
is not affine equivariant, then the breakdown at the edge is known to be zero, again see \citet{Stahel81, Hampel86}. This again
suggest that the \emph{SSCM} is not particularly robust when the underlying scatter structure is nearly singular.

\section{Appendix} \label{Appendix}
\subsection{Review of some matrix algebra} \label{matrix}
The notation used within the paper has become fairly standard and is as follows. If $\bA$ is a $p \times q$ matrix, then $vec(\bA)$ is the $pq$-dimensional vector formed by 
stacking the columns of $\bA$. If $\bA$ is a \mbox{$p \times q$} matrix and $B$ is a $r \times s$ matrix, then the Kronecker product of $A$ and $B$ is the $pr \times qs$ partitioned 
matrix $\bA \otimes \bB = [a_{jk}\bB]$. The commutation matrix $\mK_{a,b}$ is the $ab \times ab$ matrix $\sum_{i=1}^a \sum_{j=1}^b \bJ_{ij} \otimes \bJ^T_{ij}$, where $\bJ_{ij}$ is an 
$a \times b$ matrix with a one in the $(i, j)$ position and zeros elsewhere. Algebraic properties involving the $vec$ transformation, the Kronecker product and the 
commutation matrix can be found e.g.\ in \cite{Bilodeau99}. Some important properties, which are to be used without further reference, are: 
\begin{equation} \label{vec}
\begin{array}{ccc} 
vec(\bA\bB\bC) = (\bC^T \otimes \bA)vec(\bB), & \quad \mK_{d,d}^2 = \bI_{d^2}, \quad  & (\bA \otimes \bB)(\bC \otimes \bD) = (\bA\bC \otimes \bB\bD), \\[6pt]
\mK_{r,p}(\bA \otimes \bB) = (\bB \otimes \bA)\mK_{s,q}, & \quad \mbox{and} \quad & \mK_{p,q}vec(\bA) = vec(\bA^T).
\end{array}
\end{equation}

\subsection{Asymptotic Distribution of Eigenprojections  of a Scatter Estimate} \label{asy-dis}
The proof of equation (\ref{varP}) is given in this section. To begin, results for the asymptotic distribution for eigenprojections in general given in \cite{Tyler81}
are partially reviewed and extended here. Let $\bM$ be a $d \times d$ positive definite symmetric matrix with eigenvalues $\eta_{(1)} > \eta_{(2)} > \cdots > \eta_{(m)}$
having multiplicities $d_1, \ldots, d_m$ respectively, and corresponding eigenprojections $\bP_1, \ldots, \bP_m$. Suppose $\bM_{n}$ is a sequence of random positive definite 
symmetric matrices such that $\sqrt{n}(\bM_n - \bM) \rightarrow_{\mD} \bN$, with $vec(\bN)$ having a $Normal_{d^2}(\bzero,\mV_{\bM})$ distribution.
Similar to the definitions of $\bhP_{\mT,j}$ and $\bhP_{\mS,j}$, let $\bhP_j$ denote the consistent estimate of $\bP_j$. It then follows from Theorem 4.1 in \cite{Tyler81}
that
\[ \sqrt{n}(\bhP_j - \bP_j) \rightarrow_{\mD} \bP_j\bN(\bM - \eta_{(j)}\bI_{d})^+ + (\bM - \eta_{(j)}\bI_d)^+ \bN\bP_j, \]
where $\bA^+$ denotes the Moore-Penrose generalized inverse of $\bA$.
Using the properties given in (\ref{vec}), and noting that $vec\{\bN\} = vec\{\bN^T\} = \mK_{d,d}vec\{\bN\}$, one obtains
\[ \sqrt{n} ~ vec\left\{\bhP_j - \bP_j\right\} \rightarrow_{\mD} \left(\bI_{d^2} + \mK_{d,d}\right) \left\{\bP_j \otimes \left(\bM - \eta_{(j)}\bI_d\right)^+ vec\{\bN\} \right\}, \]
and hence $\sqrt{n} ~ vec\left\{\bhP_j - \bP_j\right\} \rightarrow_{\mD} Normal_{d^2} \left( \bzero, \mV_{\bP_j} \right)$ where
\begin{equation} \label{VarP}
\mV_{\bP_j} = \left(\bI_{d^2} + \mK_{d,d}\right) \left\{\bP_j \otimes \left(\bM - \eta_{(j)}\bI_d\right)^+\right\} \mV_{\bM}\left\{\bP_j \otimes \left(\bM - \eta_{(j)}\bI_d\right)^+ \right\}
 \left(\bI_{d^2} + \mK_{d,d}\right) 
\end{equation}

Suppose now that $\mV_{\bM}$ is of the form
\begin{equation} \label{affscat}
\mV_{\bM} = \sigma_1 (\bI_{d^2} + \mK_{d,d}) (\bM \otimes \bM) + \sigma_2 vec\{\bM\}vec\{\bM\}^T,
\end{equation}
as is the case when $\bM_n$ is an affine equivariant scatter estimate based on a random sample from an elliptical distribution, see \citet{Tyler83}, then
(\ref{VarP}) becomes
\begin{equation} \label{aVarP}
\mV_{\bP_j} = \sigma_1 \left(\bI_{d^2} + \mK_{d,d}\right) \sum_{k=1, k \ne j}^m \frac{\eta_{(j)}\eta_{(k)}}{(\eta_{(k)}-\eta_{(j)})^2}(\bP_j \otimes \bP_k + \bP_k \otimes \bP_j).
\end{equation}
To verify this last result note that $\bM\bP_j = \eta_{(j)}\bP_j$, $\bP_j\bP_j = \bP_j$, and for \mbox{$k \ne j$}, $\bP_j\bP_k = \bzero$. 
So, for \mbox{$k \ne j$},
$(\bP_j \otimes \bP_k)vec(\bM) = vec\left\{\bP_k \bM \bP_j \right\} = \bzero$, 
\mbox{$ (\bP_j \otimes \bP_k)(\bM \otimes \bM)(\bP_j \otimes \bP_k) = \eta_{(j)}\eta_{(k)} (\bP_j \otimes \bP_k)$,} and
\mbox($\bP_j \otimes \bP_k)\mK_{d,d}(\bM \otimes \bM)(\bP_j \otimes \bP_k) = \mK_{d,d}(\bP_k \otimes \bP_j)(\bM \otimes \bM)(\bP_j \otimes \bP_k) = \bzero.$
Hence,  
\[ \ \left(\bI_{d^2} + \mK_{d,d}\right) (\bP_j \otimes \bP_k)\mV_{\bM}(\bP_j \otimes \bP_k)\left(\bI_{d^2} + \mK_{d,d}\right) =
\sigma_1\eta_{(j)} \eta_{(j)}\left(\bI_{d^2} + \mK_{d,d}\right) (\bP_j \otimes \bP_k)\left(\bI_{d^2} + \mK_{d,d}\right). \]
Also, $\left(\bI_{d^2} + \mK_{d,d}\right) (\bP_j \otimes \bP_k)\left(\bI_{d^2} + \mK_{d,d}\right) = \left(\bI_{d^2} + \mK_{d,d}\right) (\bP_j \otimes \bP_k + \bP_j \otimes \bP_k)$, and for 
distinct $j, k, k'$, \mbox{$(\bP_j \otimes \bP_k)\mV_{\bM}(\bP_j \otimes \bP_{k'}) = \bzero$.} Expression (\ref{aVarP}) then follows by using the spectral representation 
$\left(\bM - \eta_{(j)}\bI_d\right)^+ =\sum_{k \ne j} (\eta_{(k)} - \eta_{(j)})^{-1} \bP_k$.

For Tyler's scatter estimate, the eigenprojections for $\bhT_n$ and for $\bhT_{o,n} = \left(d/\trace\{\bGamma^{-1}\bhT_n\}\right)\bhT_n$ are
the same, namely$\bhP_{\mT,j}$.  Under any elliptical distribution, the asymptotic variance for $\bhT_{o,n}$ is shown in \cite{Tyler87a} to have the form 
(\ref{affscat}) with $\sigma_1 = (d+2)/d$, $\sigma_2 = 2/d$ and with $\bM = \bGamma$. Equation (\ref{aVarP}) thus applies, and this establishes (\ref{varP}) 
for the case $\mV_{\mT,j}(\bGamma)$. 

The result (\ref{varP}) for $\mV_{\mS,j}(\bGamma)$ is more complicated to establish since the \emph{SSCM} is not affine equivariant. 
Using the spectral value decomposition \mbox{$\bGamma = \bQ\bLambda\bQ^T$}, it follows that
\mbox{$\mV_{\mS,j}(\bGamma) = (\bQ \otimes \bQ) \mV_{\mS,j}(\bLambda)(\bQ \otimes \bQ)^T$}, and so it is sufficient to consider the
the special case $\bGamma = \bLambda = diagonal(\lambda_1, \ldots, \lambda_d)$. 
The asymptotic variance of $\bhS_n$ is then given by  
\mbox{$\mV_\mS = var\left(vec\{\btheta\btheta^{T}\}\right) =  E[\btheta\btheta^{T} \otimes \btheta\btheta^{T}] - vec(\bDelta)vec(\bDelta)^{T}$},
where $\btheta = \sim_\mD ACG_d(\bLambda)$. We proceed by first finding an expression for $E[\btheta\btheta^{T} \otimes \btheta\btheta^{T}]$.

Let $\bee_1, \ldots \bee_d$ represent the Euclidean bases elements in $\mathbb{R}^{d}$, then
\[ E[\btheta\btheta^{T} \otimes \btheta\btheta^{T}] = \sum_{p=1}^d \sum_{q=1}^d \sum_{r=1}^d \sum_{s=1}^d a_{pqrs} (\bee_p \bee_q^T \otimes \bee_r \bee_s^T  ), \]
where $a_{pqrs} = E[\theta_p\theta_q\theta_r\theta_s]$. Since $\btheta$ is symmetrically distributed in each coordinate, it follows that $a_{pqrs} = 0$ unless
$p=q=r=s$, $p=q$ and $r = s$, $p=r$ and $q=s$, or $p=s$ and $q=r$. Also, since $ (\bee_p \bee_q^T \otimes \bee_q \bee_p^T) =\mK_{d,d} (\bee_p \bee_p^T \otimes \bee_q \bee_q^T)$,
$E[\btheta\btheta^{T} \otimes \btheta\btheta^{T}]$ can be expressed as
\[ (\bI_{d^2} + \mK_{d,d}) \left(\sum_{p=1}^d \sum_{q=1}^d  \gamma_{pq} (\bee_p \bee_p^T \otimes \bee_q \bee_q^T) - 
\sum_{p=1}^d \gamma_{pp} (\bee_p \bee_p^T \otimes \bee_p \bee_p^T ) \right) + \sum_{p=1}^d \sum_{q=1}^d \gamma_{pq} (\bee_p \bee_q^T \otimes \bee_p \bee_q^T), \]
where $\gamma_{pq} = E[\theta_p^2\theta_q^2]$. 

Equation (\ref{VarP}) can now be applied to $\bhP_{\mS,j}$ with $\bM = \bXi = \bDelta = diagonal(\phi_1, \ldots \phi_d)$
and $\bP_j = \bE_j = \sum_{i = m_j + 1}^{m_j + d_j} \bee_i\bee_i^T$. The term
$\left\{\bE_j \otimes \left(\bDelta - \phi_{(j)}\right)^+\right\}vec(\bDelta) = vec(\left(\bDelta - \phi_{(j)}\right)^+\bDelta\bE_j) = \bzero$, and for $p \ne q$,  
\mbox{
$\left\{\bE_j \otimes \left(\bDelta - \phi_{(j)}\bI_d\right)^+\right\}(\bee_p \bee_q^T \otimes \bee_p \bee_q^T) \left\{\bE_j \otimes \left(\bDelta - \phi_{(j)}\bI_d\right)^+ \right\} = \bzero$.}
Furthermore, using the fourth property in (\ref{vec}), it follows that
\mbox{$
\left\{\bE_j \otimes \left(\bDelta - \phi_{(j)}\bI_d\right)^+\right\}\mK_{d,d} (\bee_p \bee_p^T \otimes \bee_q \bee_q^T )
 \left\{\bE_j \otimes \left(\bDelta - \phi_{(j)}\bI_d\right)^+ \right\} = \bzero.
$}
Hence, $\mV_{\bM} = \mV_{\mS}$ in (\ref{VarP}) can be replaced by $ \sum_{p=1}^d \sum_{q=1}^d  \gamma_{pq} (\bee_p \bee_p^T \otimes \bee_q \bee_q^T). $ 

Next, observe that the term
\[
\left\{\bE_j \otimes \left(\bDelta - \phi_{(j)}\bI_d\right)^+\right\}(\bee_p \bee_p^T \otimes \bee_q \bee_q^T) \left\{\bE_j \otimes\left(\bDelta - \phi_{(j)}\bI_d\right)^+ \right\}
= \left(\phi_q-\phi_{(j)}\right)^{-2} (\bee_p \bee_p^T \otimes \bee_q \bee_q^T),
\]
for $p \in \{m_j +1, \ldots, m_j+d_j\}$ and $q \not\in \{m_j +1, \ldots, m_j+d_j\}$, and is $\bzero$ otherwise. Also, the value of $\gamma_{pq}$ is the same for all
$p \in \{m_j +1, \ldots, m_j+d_j\}$ and $q \in \{m_k +1, \ldots, m_k+d_k\}$, say $\psi_{(j,k)}$. Hence,
\[
\mV_{\mS,j}(\bLambda) = (\bI_{d^2} + \mK_{d,d}) \left\{\sum_{k=1, k \ne j}^m  \frac{\psi_{(j,k)}}{\left(\phi_q-\phi_{(j)}\right)^2} (\bE_j \otimes \bE_k)\right\} 
(\bI_{d^2} + \mK_{d,d}).
\]
This can be seen to agree with (\ref{varP}) for the special case $\bGamma = \bLambda$ after again noting that 
\[(\bI_{d^2} + \mK_{d,d})(\bE_j \otimes \bE_k) (\bI_{d^2} + \mK_{d,d}) = (\bI_{D^2} + \mK_{d,d}) (\bE_j \otimes \bE_k + \bE_k \otimes \bE_j).\] 
The form for $\psi_{(j,k)}$ given
in (\ref{varP}) is obtained by noting $\theta_p^2 \sim_\mD \lambda_p \chi^2_{1,p}/\sum_{q=1}^d \lambda_q \chi^2_{1,q}$ with $\chi^2_{1,i}$, for $i = 1, \ldots, d$ 
having independent chi-square distributions on one degree of freedom, and 
\[ \psi_{(j,k)} = (d_jd_k)^{-1}\sum_{p=m_j+1}^{m_j + d_j}\sum_{q=m_k+1}^{m_k + d_k} \gamma_{pg}.\]

\subsection{The Gauss Hypergeometric Functions.} \label{app-hyper}

\emph{Proof of statements (\ref{hyp-psi}) and (\ref{hyp-phi}).}  The expressions for $\psi_{(1,2)}$ and $\phi_{(i)}, i =1,2$ arise from the following two observations. 
First, $U = \chi^2_{(2)}/\left(\chi^2_{(1)}+\chi^2_{(2)}\right) \sim_{\mD} Beta\left(\frac{d_{2}}{2}, \frac{d_{1}}{2}\right)$, and second
$_{2}F_{1}\left(a, b;c;k\right) = B^{-1}\left(b, c-b\right)\int^{1}_{0}x^{b-1}\left(1-x\right)^{c-b-1}\left(1-kx\right)^{-a}dx$, where 
$B\left(a, b\right) = \Gamma\left(a\right)\Gamma\left(b\right)/\Gamma\left(a+b\right)$ is the Beta function, see e.g. (15.2.1) in \cite{Abramowitz92}.
The integral representation of the Gauss hypergeometric function is valid for  $\Re\left(c\right) > \Re\left(b\right) > 0$. From this, evaluation of the
following expectation is straightforward,
\[
E\left[U^{r}\left(1-U\right)^{s}\left(1-\kappa U\right)^{-t}\right] =
\frac{B\left(\frac{d_{2}+2r}{2}, \frac{d_{1}+2s}{2}\right)}{B\left(\frac{d - d_{1}}{2}, \frac{d_{1}}{2}\right)} 
~ {}_{2}F_{1}\left(t, \frac{d - d_{1}+2r}{2}; \frac{d+2s+2r}{2}; \kappa \right).
\]
Expressions (\ref{hyp-psi}) and (\ref{hyp-phi}) then follow since, for $\kappa = 1- \rho^2$,
\[ \psi_{(1,2)} = \frac{\rho^2}{d_1d_2} E\left[ \frac{U(1-U)}{\left(1-\kappa U\right)^{2}} \right], \quad 
\phi_{(1)} = \frac{1}{d_1} E\left[ \frac{1-U}{1-\kappa U} \right] \quad \mbox{and} \quad
\phi_{(2)} = \frac{\rho^2}{d_2} E\left[ \frac{U}{1-\kappa U} \right]. \] 
More details on the above derivations and on the following derivations can be found in the dissertation by \cite{Magyar12d}. \\[4pt]

\emph{Proof of statement (\ref{hyp-are}).} 
The form for the asymptotic relative efficiency can be obtained by inserting (\ref{hyp-psi}) and (\ref{hyp-phi}) into the
expressions for $\alpha_{\mT}(\rho)$ and $\alpha_{\mS}(\rho)$ given in (\ref{alpha}), and then simplifying the ratio $\alpha_{\mT}(\rho)/\alpha_{\mS}(\rho)$
using the identity (15.2.20) in \cite{Abramowitz92}, i.e.\ 
\[ \frac{1-\kappa}{\kappa} \, _{2}F_{1}\left(a, b; c; \kappa\right) - \frac{1}{\kappa} \, _{2}F_{1}\left(a - 1, b; c; \kappa\right) + 
\frac{c-b}{c} \, _{2}F_{1}\left(a, b; c+1; \kappa\right) = 0, \]
with $a = (d_{2}+2)/2, b = 1$, and $c = (d+2)/2$. \\[4pt]

\emph{Proof of statement (\ref{dim2}).} 
The form for $\phi_{(2)}$ can be obtained by using the identity (15.1.14) in \cite{Abramowitz92}, namely
${}_2F_1\left(a,\frac{1}{2}+a;2a;\kappa\right) = 2^{2a-1}\left[1+(1-\kappa)^{1/2}\right]^{1-2a}$ with $a=1$. The form for $\phi_{(1)}$
then follows since $\phi_{(1)}+\phi_{(2)} = 1$. Finally, the form for $\psi_{(1,2)}$ follows from identity 
$(1-\kappa)^{1/2} ~ {}_2F_1\left(1+a,\frac{1}{2}+a;1+2a;\kappa\right) =2^{2a}\left[1+(1-\kappa)^{1/2}\right]^{-2a}$ using $a=1$, see (15.1.13) in 
\cite{Abramowitz92}.  \\[4pt]

\emph{Proof of statement (\ref{rho-0}).} Identity (15.1.20) in \cite{Abramowitz92} states 
${}_2F_1\left(a,b;c;1)\right) = \{\Gamma(c)\Gamma(c-a-b)\}/\{\Gamma(c-a)\Gamma(c-b)\}$ provided $c > a + b$. 
Taking the limit as $\rho \rightarrow 0$ in (\ref{hyp-are}) and applying this identity gives (\ref{rho-0}) for the case $d_1 > 2$.  
The cases $d_1 =1,2$ require special treatment since for these cases $c-a-b = (d_1-2)/2 \le 0$. Here the
the identity (15.3.3) in \cite{Abramowitz92} is useful. This states that \mbox{${}_2F_1(a; b; c; \kappa) = (1-\kappa)^{c-a-b}{}_2F_1(c - a; c - b; c; \kappa)$}. 
From this it follows that the denominator ${}_2F_1\left(2,\frac{d_2+2}{2};\frac{d+2}{2};1-\rho^2\right) \rightarrow \infty$ as $\rho \rightarrow 0$ 
and hence (\ref{rho-0}) follows.

\subsection{Limiting Value of the Finite Sample Relative Efficiency.} \label{app-arn} 
Express $\bP_1 = \bQ_1\bQ_1^T$ with $\bQ_1^T\bQ_1 = \bI_{d_1}$.  The ordered eigenvalues of $\bQ_1^T\bhP_{\mT,1}\bQ_1$ then correspond to
 $\cos^2\left(\widehat{\tau}_{1,n}\right), \ldots, \cos^2\left(\widehat{\tau}_{d_1,n}\right)$. Recall 
$\sqrt{n}(\bhP_{\mT,1} -\bP_1) \rightarrow_{\mD} \sqrt{\alpha_{\mT}(\rho)}~\bZ$,
with $vec(\bZ)$ being $Normal_{d^2}(\bzero, \mM_{1,2})$. 
This implies $n~\left(I-\bQ_1^T\bhP_{\mT,1}\bQ_1 \right) = n ~\bQ_1^T\left(\bhP_{\mT,1} -\bP_1 \right)^2\bQ_1  \rightarrow_{\mD} \alpha_{\mT}(\rho)~\bW$, 
where $\bW = \bQ_1^T\bZ^2\bQ_1 = \bQ_1^T\bZ^T\bZ\bQ_1^T$. 

Since ordered eigenvalues are continuous functions of their symmetric matrix arguments, see e.g. \cite{Kato66}, 
it follows that for $i = 1, \ldots, d_1$ , $n \{1-\cos^2\left(\widehat{\tau}_{i,n}\right)\}  = n \sin^2\left(\widehat{\tau}_{i,n}\right) $
converge jointly in distibution to $\alpha_{\mT}(\rho)~\sigma_i(\bW)$, with $\sigma_1(\bW) \ge \ldots \ge \sigma_{d_1}(\bW) \ge 0$ being the ordered
eigenvalues of $\bW$. Using the delta method, one obtains $n~\left\{\sin^2\left(\widehat{\tau}_{i,n}\right) - \widehat{\tau}^2_{i,n} \right\} \rightarrow_{\mP} 0$,
and so \mbox{$nT_n  \rightarrow_{\mD} \alpha_{\mT}(\rho)~\trace(\bW)$}. An analogous argurment gives  $n\Omega_n  \rightarrow_{\mD} \alpha_{\mS}(\rho)~\trace(\bW)$.
Since $\trace(\bW)$ has a continuous distribution, the medians of $n T_n$ and $n \Omega_n$ converge to $\alpha_{\mT}(\rho)~median\left\{\trace(\bW)\right\}$
and $\alpha_{\mS}(\rho)~median\left\{\trace(\bW)\right\}$ respectively. Hence (\ref{arn}) converges to $\alpha_{\mT}(\rho)/\alpha_{\mS}(\rho)$ when using the
median. This limit would also hold when using the expected value if it could be established that both $\{nT_n~|~ n = 1, \ldots, \infty\}$ 
and $\{ n\Omega_n~|~ n = 1, \ldots, \infty \}$ are uniformly integrable. We leave this problem for possible future research. 

\bibliographystyle{plain}

\end{document}